\documentclass[aps,pra,notitlepage,twocolumn,amsmath,amssymb,showpacs,nofootinbib,superscriptaddress]{revtex4-2}
\usepackage{graphicx}
\usepackage{amsmath}
\usepackage{bm}
\usepackage{xcolor}
\usepackage{comment}
\usepackage[citecolor=black,linkcolor=black]{hyperref} 

\newcommand{\be}{\begin{equation}}  
\newcommand{\ee}{\end{equation}}
\newcommand{\ba}{\begin{array}}
\newcommand{\ea}{\end{array}}
\newcommand{\bea}{\begin{eqnarray}}
\newcommand{\eea}{\end{eqnarray}}
\newcommand{\bra}{\langle}
\newcommand{\ket}{\rangle}

\newcommand{\nn}{\nonumber}

\begin{document}

\title{Cyclic quantum engines enhanced by strong bath coupling}

\author{Camille L Latune}
\thanks{Equal contribution to this work.}
\affiliation{Univ Lyon, ENS de Lyon, CNRS, Laboratoire de Physique, F-69342 Lyon, France}

\author{Graeme Pleasance}
\thanks{Equal contribution to this work.}
\affiliation{Department of Physics, University of Stellenbosch, Stellenbosch, 7600, South Africa}

\author{Francesco Petruccione}
\affiliation{Department of Physics, University of Stellenbosch, Stellenbosch, 7600, South Africa}
\affiliation{National Institute for Theoretical and Computational Sciences (NITheCS), South Africa}
\affiliation{School of Data Science and Computational Thinking, University of Stellenbosch, Stellenbosch, 7600, South Africa}
\date{\today}
\begin{abstract}
While strong system-bath coupling produces rich and interesting phenomena, applications to quantum thermal engines have been so far pointing mainly at detrimental effects. The delicate trade-off between efficiency loss due to strong coupling and power increase due to faster equilibration, while acknowledged, remains largely unexplored owing to the challenge of assessing precisely the equilibration time. Here, we overcome this obstacle by exploiting exact numerical simulations based on the hierarchical equations of motion (HEOM) formalism. We show that a quantum Otto cycle can perform better at strong (but not ultrastrong) coupling in that the product of the efficiency and output power is maximized in this regime. In particular, we show that strong coupling allows one to obtain engines with larger efficiency than their weakly coupled counterparts, while sharing the same output power. Conversely, one can design strongly coupled engines with larger power than their weakly coupled counterparts, while sharing the same efficiency. Overall, our results provide situations where strong coupling can directly enhance the performance of thermodynamic operations, re-enforcing the importance of studying quantum thermal engines beyond standard configurations. 
\end{abstract}

\maketitle

\section{Introduction}

The formulation of quantum thermodynamics in the context of heat engines seeks to reveal the fundamental limitations on heat to work conversion at the microscopic scale \cite{Kosloff_2013,Kosloff_2014, Mitchison_2019,Latune_2021,Bhattacharjee_2021}. In conventional models of quantum heat engines, the exchange of energy between the working system and heat baths is assumed to occur under weak coupling between the two. This enables the time evolution of the system to be described in terms of a second-order master equation \cite{FrancescoBook}, which if in Gorini-Kossakowski-Sudarshan-Lindblad form \cite{Gorini_1976,Lindblad_1976}, produces a formulation consistent with macroscopic thermodynamic laws. However, while this framework has many practical uses, its restrictive assumptions limit the study of strong coupling effects that can profoundly impact the operation of quantum thermal machines, including the generation of non-thermal states, steady state coherences, and faster equilibration \cite{Smith_2014,Purkayastha_2020,Cresser_2021,PRA2022,Quanta2022}. 

Recently, the role of strong coupling in thermodynamic cycles has been examined by several authors in the context of continuous quantum thermal engines \cite{Gallego_2014,Strasberg_2016,Gelbwaser_2015,Katz_2016, Mu_2017, NazirChapter, Ivander_2021}, quantum Carnot engines \cite{Perarnau_2018}, and quantum Otto engines \cite{Pozas_2018,Wiedmann_2020, Newman_2017,Newman_2020, Camati_2020, Shirai_2021, Ptaszynski_2022}.
One conclusion common to all studies on cyclic engines is that increasing the coupling strength between the working medium and baths tends to reduce the engine's work output and efficiency \cite{Perarnau_2018,Wiedmann_2020,Camati_2020,Shirai_2021}. The reasons for this stem from the work costs needed in switching off such interactions, which are neglected under a standard weak coupling treatment, as well as the generation of coherence via the interaction with the baths \cite{Newman_2020}. On the other hand, with stronger coupling one also expects an accompanying increase in the output power due to the faster equilibration of the working medium (although not for too strong coupling where the decoupling costs outweigh the extractable work). It is therefore of interest to reveal whether strong coupling may be implemented as a ``resource" for optimizing the trade-off between power and efficiency. In particular, since a precise calculation of the power relies on determining the time interval needed for the working medium and baths to reach equilibrium, this optimization is difficult to perform in general due to requiring full non-perturbative access to the cycle dynamics. As such, a comprehensive understanding of the effect of strong coupling on the power-efficiency balance of cyclic quantum engines -- besides the detrimental impact of the decoupling costs -- has yet to be established. 

In this paper, we look to fill this gap, and argue more specifically that quantum Otto engines can actually perform best in the strong coupling regime. To overcome the difficulty in performing the optimization over the coupling strength, we employ the numerically exact HEOM method \cite{Tanimura_1989,Ishizaki_2005,Tanimura_2020,Tanimura_2006,Ishizaki_2009}, which enables accurate estimations of the equilibration time of the working medium based on calculations of the steady state. As main results we show that the maximum of both the output power and hybrid figure of merit (HFOM) accounting for the power-efficiency balance occurs for strong system-bath coupling. In parallel, we investigate the performance of an {\it interrupted} quantum Otto engine where one of the isochoric strokes (see Fig. \ref{fig:1}) is stopped before thermalization occurs. For such cases the output power is further influenced by two competing effects: on the one hand, one expects a general reduction in power due to less heat being exchanged with each of the baths, while on the other hand it is increased by the faster cycle completion \cite{Newman_2020,Quadeer_2021}. This follows related studies investigating the role of non-Markovian effects on the performance of a quantum Otto engine \cite{Shirai_2021} and refrigerator \cite{Camati_2020}, where an increase in power due to interrupted isochores was observed.

Here, we also observe a significant increase in the output power and HFOM for an interrupted cycle at strong to moderate bath coupling, but we note that part of the power enhancement is due to an ``over-equilibration''  phenomenon; namely, where the system populations temporarily exceed their steady state value before converging to it. As a consequence, we further show that strong coupling can be used to enhance the efficiency at fixed output power, and vice versa.

The remainder of this paper is structured as follows. In Sec. \ref{sec:2}, we introduce the strong coupling Otto cycle and outline how the corresponding quantities of interest -- including the work and heat per cycle -- are extracted from the HEOM. We then present results demonstrating the maximum of the power and HFOM to lie in the strong coupling regime. In Sec. \ref{sec:3}, we consider extensions to interrupted cycles and present corresponding calculations of the work, output power and HFOM. 
Following these results we show how strong coupling effects may enhance the performance of the interrupted engine at either fixed output power or efficiency. Finally, we outline our conclusions and discuss future developments in Sec. \ref{sec:4}.

\section{Strong coupling cycle}\label{sec:2}
\begin{figure}[t!]
\centering
\includegraphics[scale=0.38]{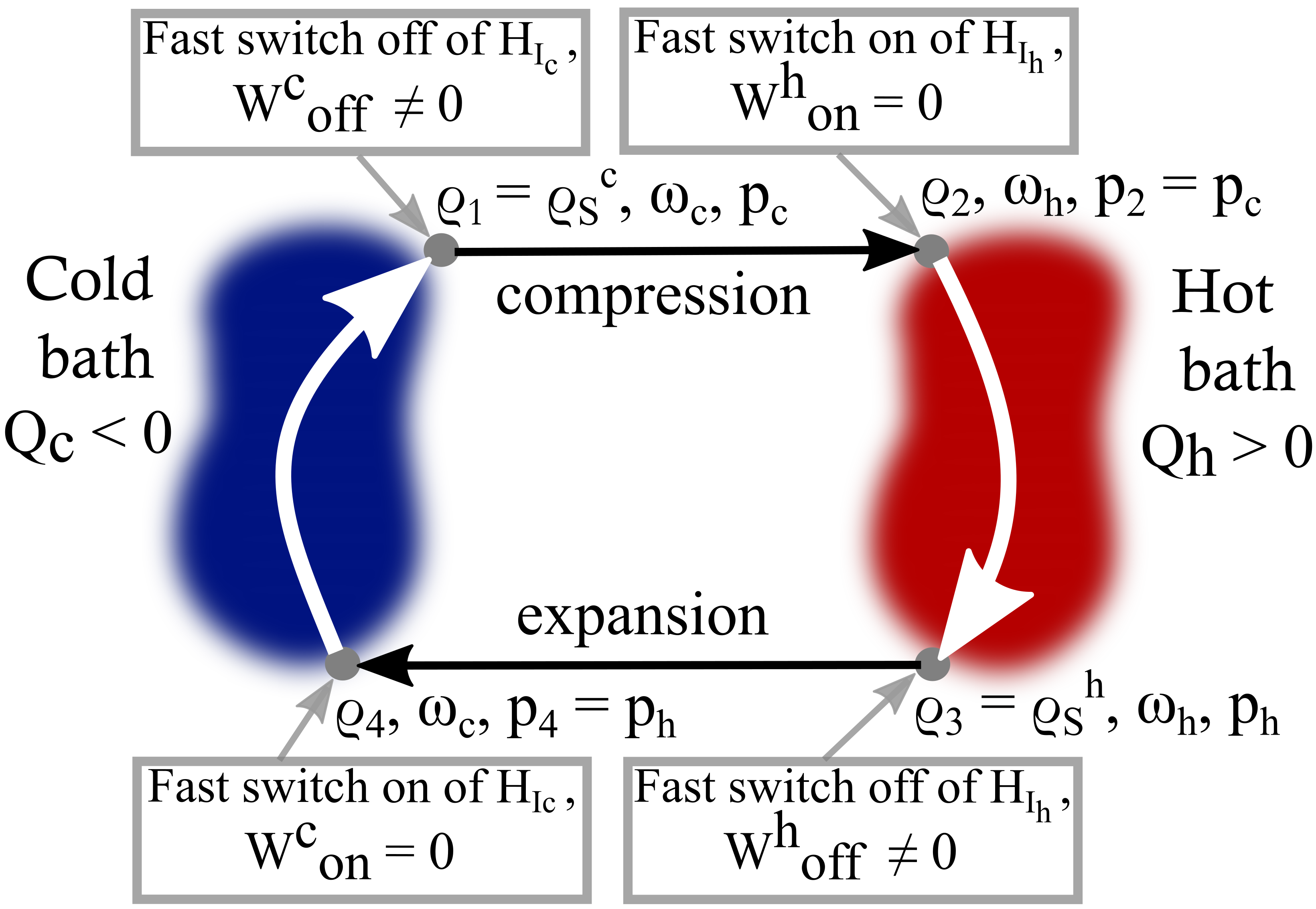}
\caption{Strong coupling Otto cycle: the two unitary strokes are represented by black arrows, while the two white arrows represent the interaction with the baths. Here, these arrows are thick to represent strong interaction with the baths, main difference with standard Otto cycles. 
Section \ref{sec:2} of the paper focuses on equilibrating isochores, where the duration of the isochoric strokes is such that the working medium reaches a steady state. This assumption is relaxed in Sec. \ref{sec:3}.}
\label{fig:1}
\end{figure}

We first consider a quantum Otto cycle \cite{Quan_2007} with the usual configuration consisting of two isochoric strokes and two unitary (adiabatic) strokes, as depicted in Fig. \ref{fig:1}, where the working medium $S$ is a two-level system (TLS). The full microscopic Hamiltonian of the model reads (we use units of $\hbar=k_B=1$ throughout)
\be
H(t) = H_S(t) + \sum_j\big(H_{B_j} + H_{I_j}\big).
\ee
Here, $H_S(t)$ is the time-dependent Hamiltonian of the TLS, while
\begin{align}
    H_{B_j}&=\sum_k\omega_kb^{\dagger}_{k_j}b_{k_j}, \\
    H_{I_j}&=V_j\sum_kg_{k_j}(b_{k_j}+b^{\dagger}_{k_j}),
\end{align}
denote the Hamiltonians of the hot/cold baths $B_j$ and interaction ($j=h,c$), with $V_j=V^{\dagger}_j$ the system coupling operator to the $j$th bath. Furthermore, $b_{k_j}$ ($b^{\dagger}_{k_j}$) and $\omega_k$ are the bosonic annihilation (creation) operator and frequency for the $k$th mode of the $j$th bath, respectively, and $g_{k_j}$ are the associated coupling strengths. Note that the interaction Hamiltonian $H_{I_j}$ contains an implicit time dependence and is only switched on during the relevant isochore. 
At the beginning of each isochore we assume the state of the system and bath to factorize as $\rho_{SB_j}(0)=\rho_S(0)\otimes\rho^{\rm th}_{B_j}$, with each bath initialized in a thermal state $\rho^{\rm th}_{B_j}=e^{-\beta_jH_{B_j}}/{\rm Tr}[e^{-\beta_jH_{B_j}}]$ at inverse temperature $\beta_j=1/T_j$. Then, the effect of the $j$th bath on the system is fully determined by the bath correlation function
\be\label{eq:TTCF}
	C_j(t)=\frac{1}{\pi}\int^{\infty}_0\hspace{-0.15cm}d\omega\,J_j(\omega)\Big[\coth\left(\frac{\beta_j\omega}{2}\right)\cos(\omega t) - i\sin(\omega t)\Big],
\ee
where $J_j(\omega)=\pi\sum_kg^2_{k_j}\delta(\omega-\omega_k)$ is the spectral density \cite{FrancescoBook}. 

For conventional Otto cycles in which the system-reservoir coupling is assumed to be vanishingly weak, the effect of each isochore is to drive the system into a thermal state at the respective bath temperature $T_j$. Here, as in Refs. \cite{Pozas_2018,Wiedmann_2020, Newman_2017,Newman_2020, Camati_2020, Shirai_2021}, we instead assume an arbitrary coupling strength between $S$ and $B_j$. This implies the steady states of the isochores to no longer be thermal states, in addition to the fact that switching on and off the interaction induces a non-negligible energetic cost to the work \cite{Newman_2017,Perarnau_2018}. While the system steady states for strong bath coupling may be approximated using recently introduced techniques \cite{Gelzinis_2020,Cresser_2021, Trushechkin_2021, PRA2022, Timofeev_2022, Quanta2022} related to the mean force Gibbs state \cite{ReviewTrushechkin}, it is useful to simulate the full cycle to obtain precisely the equilibration time. It will be also useful in the second part of the paper when analyzing the benefits of interrupting the isochores before thermalization (see Sec. \ref{sec:3}). 

\subsection{Cycle analysis}
 
{\it First stroke--}. The first stroke starts with $S$ in the state $\rho_1:=\rho^c_S$, resulting from the potentially strong interaction with the cold bath $B_c$. It is assumed that the duration of the interaction with the cold bath is long enough so that $S$ has reached its steady state. In the following we consider the TLS Hamiltonian to be of the form
\be\label{eq:H_S}
H_S(t) = \omega(t)|+\rangle\langle +| := \frac{\omega(t)}{2}(r_x\sigma_x + r_z\sigma_z + \mathbb{I}),
\ee
where $\omega(t)$ is an external control parameter, and $r_x$, $r_z$ are real numbers fulfilling $r_x^2+r^2_z=1$. The ground state of the TLS is denoted by $|-\ket$ and is associated with zero energy. The first stroke then consists of a unitary drive taking the system Hamiltonian from $H_c=\omega_c|+\ket\bra +|$ to $H_h:=\omega_h |+\ket\bra +|$ ($\omega_h>\omega_c$) in a time interval $\tau_1$; the simplest drive corresponds to $\omega(0) = \omega _c $ and $\omega (\tau_1) = \omega _h$, leading to a transformation of the form $U_{\rm comp} = e^{-i \theta(0,\tau_1)} |+\ket\bra +| + |-\ket\bra -|$, with $\theta(0,\tau_1) = \int_0^{\tau_1}du\,\omega(u)$. The state at the end of the compression stroke is therefore given as $\rho_2:= U_{\rm comp}\rho_1U_{\rm comp}^{\dag}$, which has the same populations as $\rho_1$ (in the basis $\{|+\ket,|-\ket\}$). The work injected during this stroke is\footnote{The presence of the term proportional to the identity in Eq. (\ref{eq:H_S}) is only relevant for computing the energy difference during the unitary strokes. It can be ignored for the dissipative strokes since it does not affect the dynamics neither the energy balance. This is equivalent to setting the origin of energy to zero instead of $-\omega_{c}/2$ or $-\omega_h/2$. This is an important detail, otherwise the origin of the energy is implicitly redefined each time the Hamiltonian $H_S$ is re-scaled. In particular, if one defines $H_c= \frac{\omega_c}{2} (|+\ket \bra +| -|-\ket\bra -|)$ and $H_h= \frac{\omega_h}{2} (|+\ket \bra +| -|-\ket\bra -|)$, the compression stroke becomes the one during which work is actually extracted while the expansion stroke corresponds to injection of work.} 
\be\label{eq:W_1}
W_1 := {\rm Tr} [\rho_2H_h-\rho^c_SH_c] = p_{c}(\omega_h -\omega_c)\geq0,
\ee
with $p_j=\langle +|\rho^j_S|+\rangle$ ($j=h,c$) the excited system population after the interaction with the hot or cold bath.

{\it Second stroke--}. This is an isochore where $S$ is put in contact with the hot bath $B_h$. The difference from a usual isochore is that we do not make any assumption about the strength of the coupling between $S$ and $B_h$. Consequently, the energy contributions related to switching on and off the system-bath interaction $H_{I_j}$ cannot be neglected. For simplicity, we consider an instantaneous switching on/off of the bath interaction taking place on a timescale smaller than the evolution timescale of the composite system $S+B_j$, so that $\rho_{SB_j}$ is assumed to be unchanged during the switching steps. This implies that there is no heat exchange and the work contribution is 
\bea
W_{\rm on}^h &=& \int_{\tau_1}^{\tau_1+\tau_{\rm on}}dt \,{\rm Tr}\left[\rho_2\rho_{B_h}^{\rm th}\frac{d}{dt}H_{I_h}(t) \right]\nn\\
             &=& {\rm Tr}_S[\rho_2H_{I_h}]\sum_kg_{k_j}{\rm Tr}_{B_h}[\rho_{B_h}^{\rm th}(b_{k_h}+b^{\dagger}_{k_h})]=0. \hspace{0.5cm}
\eea
As indicated above, we first assume that the duration of the isochores are long enough so that $S$ reaches the steady state. The duration of the hot isochore is then equal to the equilibration time of the TLS in contact with the hot bath, $\tau^{\rm eq}_h$, with $\rho^h_S:=\rho_S(\tau^{\rm eq}_h)$. The associated heat transfer -- the energy supplied by $B_h$ -- is given as \cite{Kato_2016,Esposito_2010,Rivas_2020}
\bea\label{eq:Q_h}
 Q_h &:= -&{\rm Tr}\left[(\rho_{SB_h}^{\rm ss} - \rho_2\rho_{B_h}^{\rm th})H_{B_h}\right] \nn\\
 &=&  {\rm Tr}[(\rho^h_S - \rho_2)H_h] + {\rm Tr}[\rho_{SB_h}^{\rm ss}H_{I_h}]\nn\\
  &=&  \omega_h(p_h-p_c) + {\rm Tr}[\rho_{SB_h}^{\rm ss}H_{I_h}]
\eea
where we have used the total energy conservation $ {\rm Tr} [(\rho_{SB_h}^{\rm ss}-\rho_2\rho_{B_h}^{\rm th})(H_h+H_{I_h}+H_{B_h}) ]=0$ in the second line, and where $\rho_{SB_h}^{\rm ss}$ denotes the steady state of $S+B_h$. The work associated with decoupling $S$ from $B_h$ is (also assuming instantaneous switching-off)
\bea\label{eq:W_h_off}
W_{\rm off}^h &=&\int_{\tau_1+\tau_h}^{\tau_1+\tau_h+\tau_{\rm off}}dt\,{\rm Tr}\left[\rho_{SB_h}^{\rm th}\frac{d}{dt}H_{I_h}(t)\right] \nn\\
&=& -{\rm Tr}[\rho_{SB_h}^{\rm ss}H_{I_h}].
\eea 
One interesting question is the sign of $W_{\rm off}^h$, or in other words, whether one spends or extracts work during the switching off. We show in Appendix \ref{appupperbound} that the interaction energy ${\rm Tr}[\rho_{SB_j}H_{I_j}]$ is always negative in the steady state, ${\rm Tr}[\rho^{\rm ss}_{SB_j}H_{I_j}]<0$. Note, however, that if the initial state of the stroke is out of equilibrium, it may also be negative at intermediate times.  

{\it Third stroke--}. In this second isentropic stroke, corresponding to the expansion stroke, the system Hamiltonian $H_h$ is unitarily brought back to $H_c$ in a time interval $\tau_3$. As in the first stroke, the simplest drivings leads to a transformation of the form $U_{\rm exp} = e^{-i \theta_{\rm exp}(\tau_1+\tau_h,\tau_1+\tau_h+\tau_2)} |+\ket\bra +| + |-\ket\bra -|$, where $\theta_{\rm exp}(\tau_1+\tau_h,\tau_1+\tau_h+\tau_2)$ is the phase accumulated during the driving and depends on the detail of the driving protocol. The final state is $\rho_4 = U_{\rm exp}\rho_3 U_{\rm exp}^{\dag}$, and the extracted work is
\be\label{eq:W_3}
 W_3 := {\rm Tr}[\rho_4 H_c - \rho^h_S H_h] = p_h(\omega_c-\omega_h) \leq 0.
\ee

{\it Fourth stroke--}. The final stroke comprises a cold isochore, whereby the system interacts with the cold bath over a time interval $\tau^{\rm eq}_c$ (equilibration time of the TLS and cold bath). Since $S$ and $B_c$ are assumed to reach equilibrium the system is returned to the initial state of the cycle $\rho_1=\rho^c_S$ at the end of the interaction. Again, the work cost associated with switching on the coupling is null, while the switching off work cost is 
\be\label{eq:W_c_off}
W_{\rm off}^c = -{\rm Tr}[\rho_{SB_c}^\text{ss}H_{I_c}].
\ee
Lastly, the heat exchanged with the cold bath reads
\bea\label{eq:Q_c}
Q_c&:=-&{\rm Tr}\left[(\rho_{SB_c}^{\rm ss} - \rho_4\rho_{B_c}^{\rm th})H_{B_c}\right] \nn\\
    &=& \omega_c (p_c-p_h) +{\rm Tr}[\rho_{SB_c}^\text{ss} H_{I_c}].
\eea

Now that we discussed the cycle in detail we can proceed with introducing the relevant performance measures. For a thermal machine operating as an engine, one must have a positive extracted work 
\be\label{eq:W_ext}
W_{\rm ext}: = -W_1 -W_{\rm off}^h -W_3 - W_{\rm off}^c,
\ee
which includes the usual work contributions from the compression and expansion (adiabatic) strokes $W_1$ and $W_3$, as well as the contributions from switching off the bath couplings. The corresponding efficiency is
\bea
\eta &:=& \frac{W_{\rm ext}}{Q_h}=1 + \frac{Q_c}{Q_h}  \nn\\
&=& 1-\frac{\omega_c(p_h-p_c)  +W_{\rm off}^c}{\omega_h(p_h-p_c) - W_{\rm off}^h}\nn\\
&=& \eta_{\rm Otto} - \frac{\omega_c}{\omega_h}\frac{W_{\rm off}^c/\omega_c + W_{\rm off}^h /\omega_h}{p_h-p_c - W_{\rm off}^h/\omega_h},
\eea
where 
\be\label{eq:eff_otto}
\eta_{\rm Otto}:= 1 -\frac{\omega_c}{\omega_h}
\ee
is the Otto efficiency. From the above expression, one can see directly the detrimental impact of the work required to switch off the coupling with the baths. In particular, the only way to surpass the Otto efficiency (\ref{eq:eff_otto}) is to have  $W_{\rm off}^c/\omega_c + W_{\rm off}^h /\omega_h<0$, meaning that work is extracted during the switching off operations. This is possible only for interrupted isochores since, as commented above, extracting work from switching off the bath coupling can happen only at intermediary times, before reaching the steady state.
The output power is 
\be\label{eq:power}
{\cal P} = \frac{W_{\rm ext}}{\tau_{\rm cyc}} \simeq \frac{W_{\rm ext}}{\tau^{\rm eq}_h + \tau^{\rm eq}_c},
\ee
where $\tau_{\rm cyc} = \tau_1 + \tau^{\rm eq}_h + \tau_2 + \tau^{\rm eq}_c$ is duration of the whole cycle. Here it is feasible to set $\tau_{\rm cyc}\simeq \tau^{\rm eq}_h + \tau^{\rm eq}_c$ since, due to the eigenstates of $H_S(t)$ being time independent, the duration of the unitary strokes can be made arbitrarily short ($\tau_{1,2}\rightarrow0$). 

One of the main challenges of studying quantum thermal machines at strong coupling is assessing the equilibration time of the isochores, which is of major importance in estimating the output power (\ref{eq:power}). To overcome this difficulty we choose to simulate the full cycle using the numerically exact HEOM method \cite{Tanimura_1989,Ishizaki_2005,Tanimura_2020}. In this approach, the interaction between the system and $j$th bath is described by a coupled set of equations of motion for the reduced system density matrix $\rho_S(t)$ and auxiliary density operators (ADOs) $\hat{\rho}_{\vec{n}_j}(t)$, given that the bath correlation function $C_j(t)$ may be expanded as a finite sum of complex exponentials $C_j(t)\simeq\sum^{K_j}_{k=0}c_{jk}e^{-iz_{jk}t} + \Delta_j\delta(t)$. Each ADO is labelled by a multi-index column vector $\vec{n}_j$ of $K_j$ non-negatives integers with $\rho_S(t)=\rho_{(0,...,0)}$. The HEOM for the ADOs can be written in the time-local form \cite{Tanimura_2006}
\be\label{eq:heom}
\dot{\hat{\rho}}_{\vec{n}_j} = -i\mathcal{L}_S(t)\hat{\rho}_{\vec{n}_j} + \sum_{\vec{n}'_j}\mathcal{D}_{\vec{n}_j,\vec{n}'_j}\hat{\rho}_{\vec{n}'_j},
\ee
where $\mathcal{L}_S(t)$ is the system Liouvillian, and $\mathcal{D}_{\vec{n}_j,\vec{n}'_j}$ is a tensor-like superoperator encoding the dissipative effects of the bath on the system. Note that the latter explicitly depends on the expansion coefficients $c_{jk}$ and exponents $z_{jk}$ of the bath correlations $C_j(t)$ and (anti-)commutators of $\hat{\rho}_{\vec{n}'_j}$ with the coupling operator $V_j$. 

By solving the HEOM we are able to not only accurately estimate the equilibration times of the isochores, but also extract all quantities of interest such as the work, heat, e.t.c., directly from the ADOs returned by Eq. (\ref{eq:heom}). In particular, the interaction terms $\text{Tr}[\rho_{SB_j}H_{I_j}]$ can be evaluated from the first-tier ADOs via \cite{Zhu_2012,Kato_2016,Song_2017}
\be
    \text{Tr}[\rho_{SB_j}H_{I_j}] = \sum_{\text{1}^{\text{st}}\text{-tier}}\text{Tr}[V_j\hat{\rho}_{\vec{n}_j}],
\ee
such that the work $W_{\text{ext}}$ and heat $Q_h$ may be determined as
\bea
W_{\text{ext}} &=& -W_1 - W_3 + \sum_{j=h,c}\sum_{\text{1}^{\text{st}}\text{-tier}}\text{Tr}[V_j\hat{\rho}_{\vec{n}_j}], \\
Q_h &=& \omega_h(p_h - p_c) + \sum_{\text{1}^{\text{st}}\text{-tier}}\text{Tr}[V_h\hat{\rho}_{\vec{n}_h}].
\eea
Further details on the implementation of the HEOM are included in Appendix \ref{appHEOM}.

\begin{figure}[t!]
\centering
\includegraphics[scale=0.57]{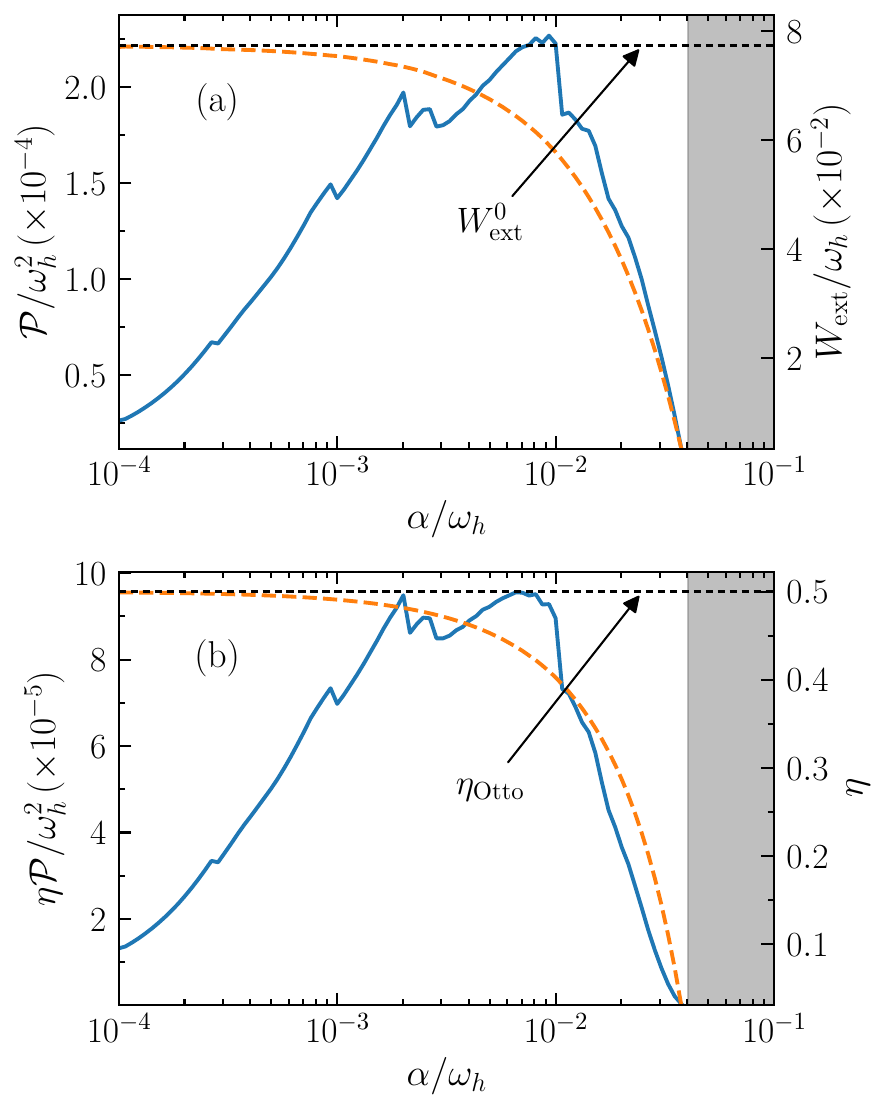}
\caption{Strong coupling cycle: (a) Power output $\mathcal{P}$ (blue solid line) and work extracted per cycle $W_{\text{ext}}$ (orange dashed line) against the coupling strength $\alpha$, taken to be equal for both baths. (b) Hybrid figure of merit (HFOM) $\eta\mathcal{P}$ (blue solid line) and efficiency $\eta$ (orange dashed line) for the same parameters. The black dashed lines correspond to the work output $W^0_{\text{ext}}$ and efficiency $\eta_{\rm Otto}$ of the equivalent weak coupling cycle. The filled regions indicate the coupling strengths for which the net work output is negative. }
\label{fig:2}
\end{figure}

\subsection{Results}
By applying the method described above, we present results for cases in which $r_z=1/2$ and $r_x=\sqrt{3}/2$. For simplicity we also assume the interaction between the TLS and each bath to be along the same $z$-direction, $V_j=\sigma_z$, and consider the bath spectral densities to be in an underdamped Brownian oscillator form \cite{Garg_1985}
\bea\label{eq:SD}
J_j(\omega) = \frac{\alpha_j\Gamma_j\omega^2_{0,j}\omega}{(\omega^2_{0,j}-\omega^2)^2+\Gamma^2_j\omega^2}.
\eea
Here, $\alpha_j$ measures the coupling strength between the system and the $j$th bath, $\omega_{0,j}$ is the location of the maximum of $J_j(\omega)$, and $\Gamma_j$ is the spectral width. For the simulation results below we integrate the hierarchy (\ref{eq:heom}) using the HEOM solver of the Python package QuTiP \cite{Johansson_2013,Lambert_2023}, with $c_{jk}$ and $z_{jk}$ obtained from the Matsubara decomposition of the bath correlation functions. In Appendix \ref{app:numeqtime} we outline how the equilibration time of the isochores is determined by integrating the HEOM until convergence to the steady state $\rho^j_S$ is reached at time $t=\tau^{\text{eq}}_j$. We also opt for resonant bath conditions throughout, whereby $\omega_{0,j} = \omega_{j}$, $\omega_c=0.5\omega_h$, $\Gamma_h = \Gamma_c = 0.05\omega_h$, $\beta_h\omega_h = 0.5$, and $\beta_c\omega_h = 2.5$. 

Figure \ref{fig:2} presents plots of the extracted work per cycle $W_\text{ext}$, power ${\cal P}$, efficiency $\eta$, and HFOM $\eta{\cal P}$ against the coupling strength $\alpha$, taken to be the same for each of the baths ($\alpha_h=\alpha_c$). For the finite time cycle outlined above the HFOM measures the trade-off between the power and efficiency and hence gauges the overall performance of the engine \cite{Shiraishi_2016, Dann_2020}. The horizontal dashed lines represent the work output 
\be
W^0_{\rm ext}:=(\omega_h-\omega_c)(p^{\rm th}_h-p^{\rm th}_c)
\ee
and efficiency (\ref{eq:eff_otto}) achieved for a standard Otto cycle at vanishing coupling $\alpha/\omega_h \rightarrow 0$, where 
\be
p^{\rm th}_j=\frac{e^{-\beta_j\omega_j}}{1+e^{-\beta_j\omega_j}}
\ee
are the excited thermal populations. One recovers -- as already observed in Refs. \cite{Pozas_2018,Wiedmann_2020, Newman_2017,Newman_2020, Camati_2020, Shirai_2021} -- that the work and efficiency decrease monotonically with the coupling strength until the bath decoupling costs $W^h_{\rm off}+W^c_{\rm off}$ outweigh the positive work contribution $-(W_1+W_3)$ from the adiabatic strokes. Importantly, we can also deduce from the same figure that the transition between the weak and finite coupling regimes occurs around $\alpha= 10^{-4}\omega_h$, where the work output and efficiency of the cycle become well approximated by $W^0_{\rm ext}$ and $\eta_{\rm Otto}$. Then, the main implication of Fig. \ref{fig:2} is that the power and corresponding HFOM are maximal around $\alpha = 10^{-2}\omega_h$, which occurs at strong coupling. While it has been noted that the power output converges to zero in the vanishing and ultrastrong coupling limits \cite{Newman_2017}, we are here able to affirm that this maximum is located well into the strong coupling regime.

It is worth highlighting that the non-smooth behavior of $\mathcal{P}$ and $\eta\mathcal{P}$ can be attributed to complementary changes in the equilibration time of the isochores when varying the system-bath coupling. In general, the equilibration time of each isochore $\tau^{\rm eq}_j$ depends not only on the strength of the coupling $\alpha$, but also on the steady state of the preceding isochore, which itself depends on $\alpha$. Hence, there is a complicated interplay between how these effects determine $\tau^{\text{eq}}_j$: on the one hand, for the cold isochore we observe the equilibration time to scale approximately as $\tau^{\text{eq}}_c\sim O(1/\alpha)$, except with isolated ``jumps'' occurring at $\alpha=10^{-2}\omega_h $ and $\alpha=2\times10^{-3}\omega_h$ -- see Appendix \ref{app:numeqtime}. In Fig. \ref{fig:2} this is what produces a sharp change in the power and HFOM at these values. On the other hand, for the hot isochore the equilibration time $\tau^{\rm eq}_h$ is much more sensitive to variations in $\alpha$. In particular, we find the jumps to be more frequent and occur at regular intervals in the coupling (again see Appendix \ref{app:numeqtime}), which result in non-smooth changes in the power at values $\alpha=10^{-2}\omega_h$, $\alpha=10^{-3}\omega_h$, and $\alpha=2\times 10^{-4}\omega_h$.

\begin{figure}
\centering
\includegraphics[scale=0.54]{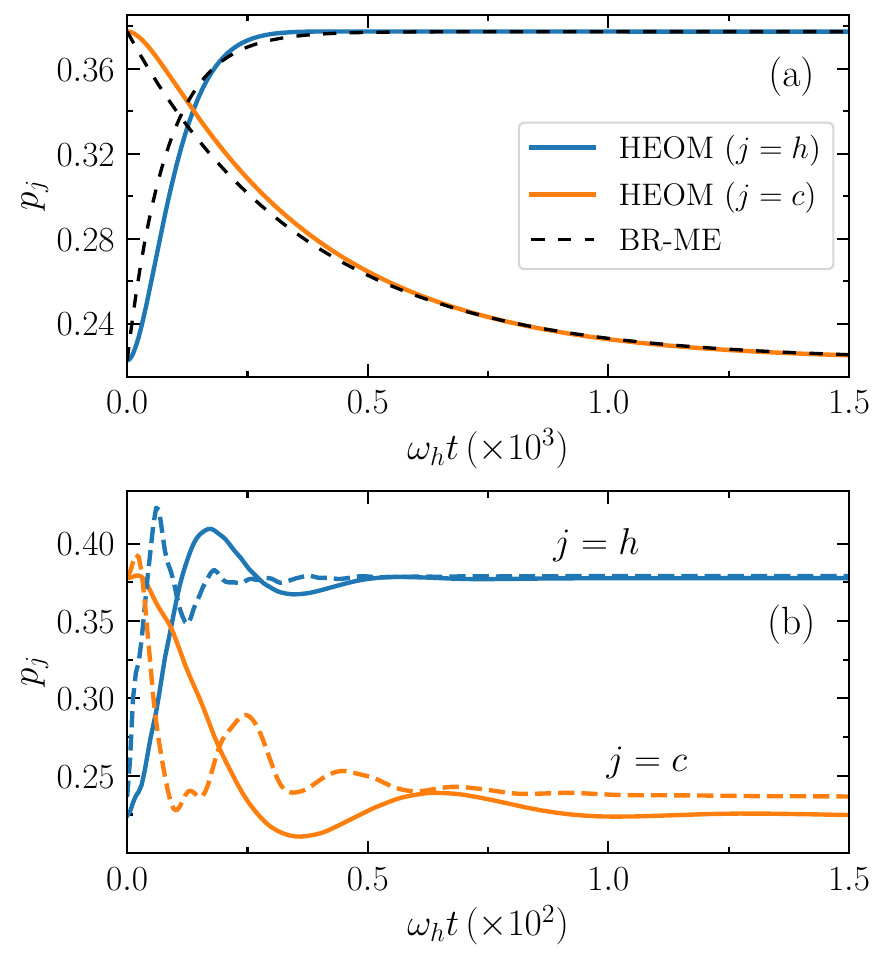}
\caption{(a) Excited state populations along the hot and cold isochores ($j=h,c$) obtained from the HEOM (solid lines) and the BR-ME (dashed lines) for $\alpha=10^{-4}\omega_h$. (b) The same populations obtained from the HEOM for coupling strengths $\alpha=0.01\omega_h$ (solid lines) and $\alpha=0.1\omega_h$ (dashed lines). }
\label{fig:3}
\end{figure}

\begin{figure*}[t!]
\centering
(a)\includegraphics[scale=0.36]{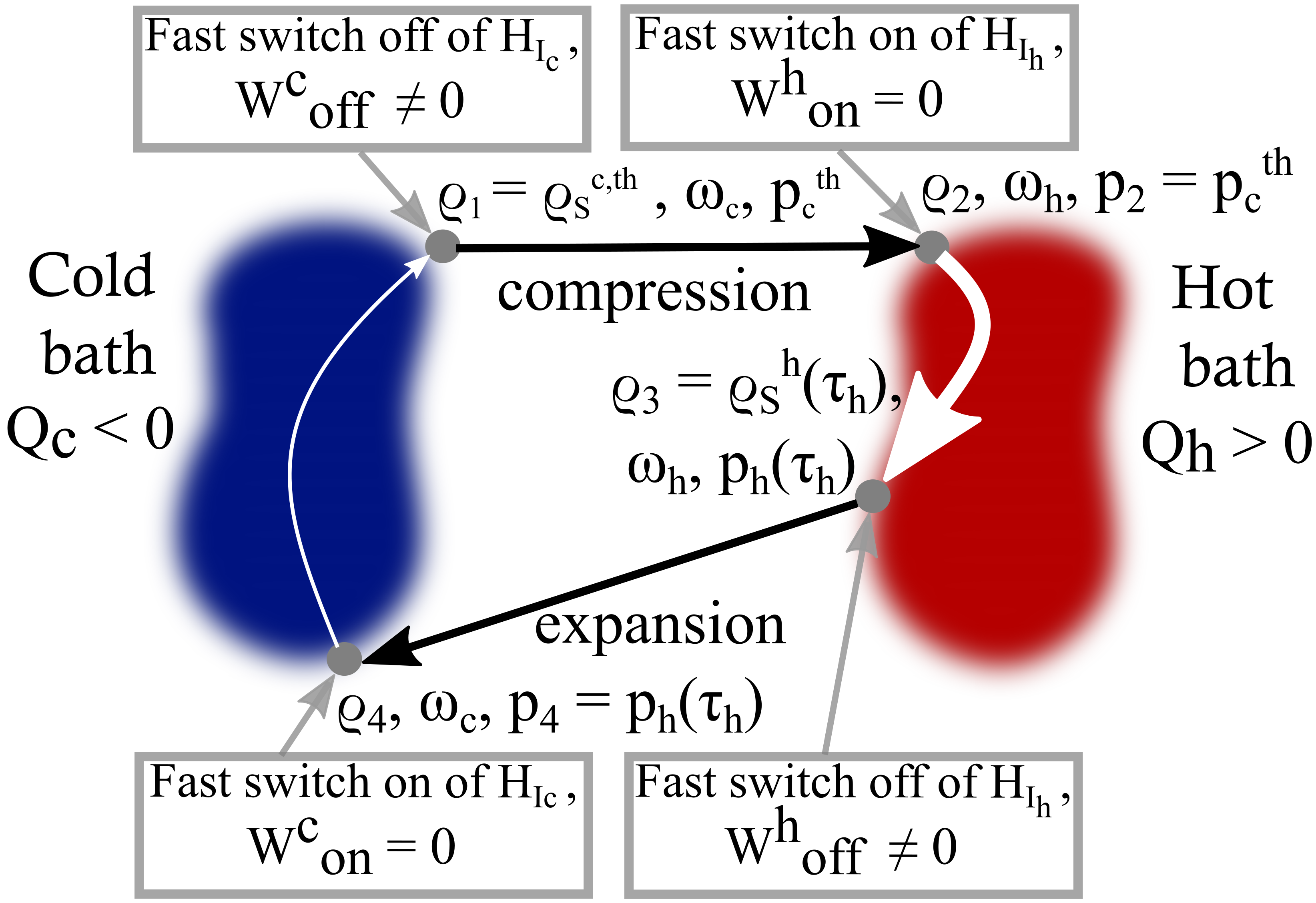}\hspace{-0.3cm}~~~(b)\includegraphics[scale=0.36]{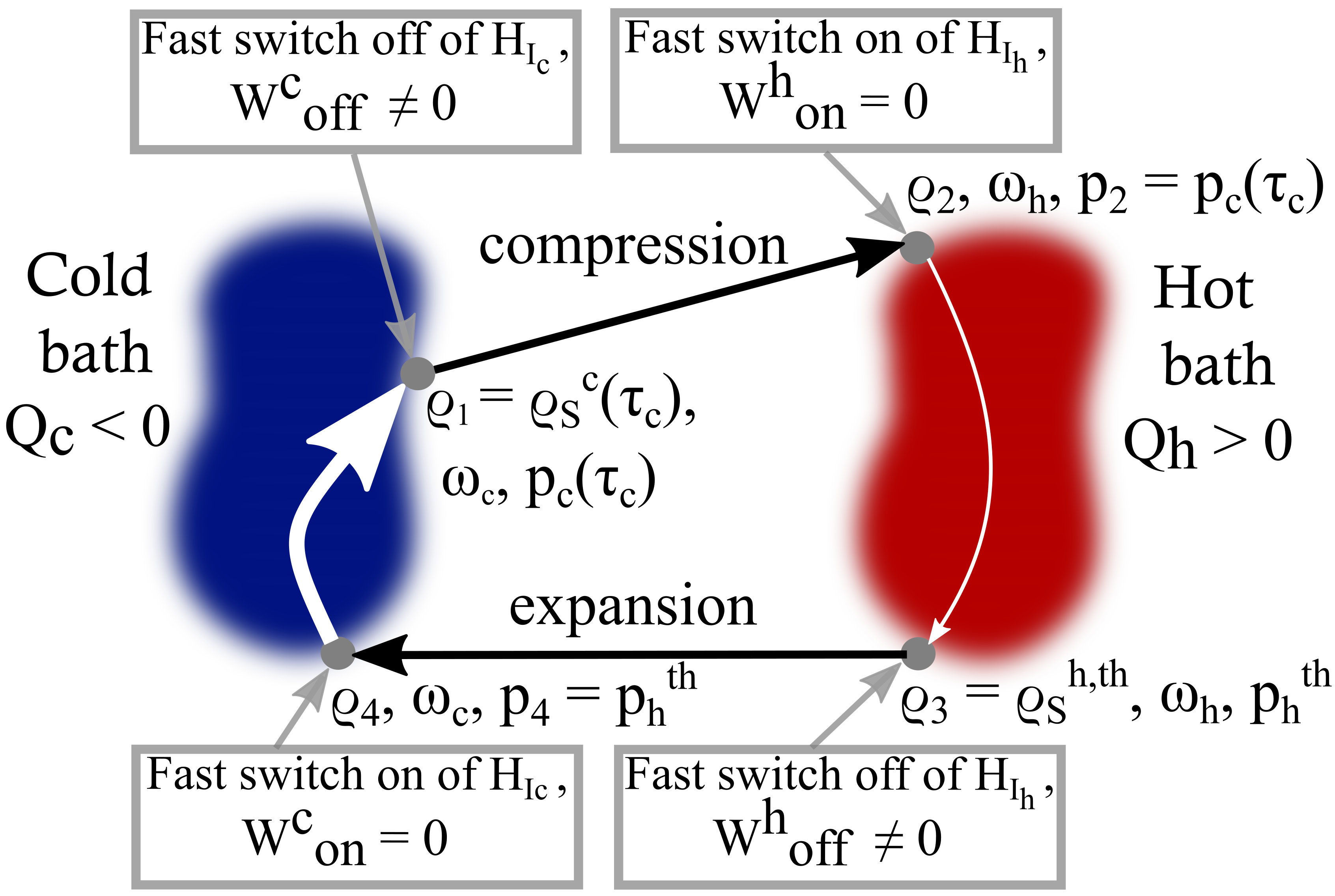}
\caption{(a) Otto cycle with interrupted hot isochore and (b) with interrupted cold isochore. The two unitary strokes are represented by black arrows, and the white arrows stand for the interaction with the baths. A thick arrow represents a strong coupling, a thin one represents a vanishing coupling, and a short one corresponds to an interaction interrupted before reaching equilibrium. }
\label{fig:4}
\end{figure*}

In Fig. \ref{fig:3}(a), to further illustrate the transition between the weak and finite (strong) coupling regimes the excited state populations for $\alpha=10^{-4}\omega_h$ are compared with those simulated using a second-order Bloch-Redfield master equation (BR-ME) \cite{Redfield_1965}, which treats the reduced system dynamics under the Born-Markov approximations. Despite there being discrepancies at short times the predictions of the BR-ME are seen to converge towards the HEOM result in the steady state. Hence, this is consistent with the conclusion drawn from Fig. \ref{fig:2} that the model describes a standard weak coupling cycle for $\alpha\sim10^{-4}\omega_h$. 

Considering stronger coupling in Fig. \ref{fig:3}(b), the same populations are plotted against the time duration of the isochores for $\alpha = 0.01 \omega_h$ and $\alpha = 0.1\omega_h$. The first observation here is that equilibration happens on a much shorter timescale than at weak coupling (almost one order of magnitude difference). Secondly, for both isochores the excited population $p_j$ temporarily exceeds its steady state value for $t<\tau^{\rm eq}_j$, which is a distinct feature of strong coupling. For the hot isochore this corresponds to the system being ``over-heated" before reaching equilibrium. A similar phenomenon happens for the cold isochore, but this time the excited population reaches values below that in the steady state; hence the system is ``over-cooled". We refer to both these phenomena as ``over-equilibration". Importantly, since the work extracted from the cycle depends on the population difference at the start of the adiabatic strokes [see Eqs. (\ref{eq:W_1}), (\ref{eq:W_3}) and (\ref{eq:W_ext})], one may then ask whether interrupting the isochores before equilibration can lead to an improvement in the engine performance. This is what we shall focus on examining for the remainder of the paper. 

\section{Interrupted isochores}\label{sec:3}
We now turn to investigate quantum Otto cycles with isochores {\it interrupted} before $S$ reaches a steady state. As commented in the introduction, one expects a trade-off between power increase stemming from a reduction of the cycle duration versus a power decrease due to a reduction of the heat exchanges (and therefore of the extracted work). On the other hand, as commented in the end of the last section, the phenomenon of over-equilibration [Fig. \ref{fig:3}(b)] has the potential to bring interesting effects for interrupted isochores.

There are two major challenges related to cycles composed of interrupted isochores. Firstly, the state of $S$ at an arbitrary instant of time $t$ depends on the state of $S$ at the beginning of the isochore. Thus, when finalizing the first cycle, the state of $S$ is no longer the equilibrium state $\rho^c_S$ since the cold isochore has been interrupted before reaching equilibrium. This same state depends on the density matrix $\rho_4$ at the beginning of the cold isochore, which depends itself of the state at the beginning of the hot isochore, $\rho_2$, and thus ultimately on the state in which we initialized the cycle $\rho_0$. In order to close the cycle (final state equal to initial state), we need to find states which are invariant under a cycle.
While this would be analytically tractable for weak coupling, it is a very challenging task for arbitrary coupling strengths. 
One simple solution is to build a cycle with only one interrupted isochore, while the second one is an equilibrating isochore yielding a steady state independent of the initial conditions, thus guaranteeing a closed cycle. In the following we enact this setup by considering two ``complementary'' cycles: the first one composed of a cold equilibrating isochore and an interrupted hot isochore, and the second composed of an interrupted cold isochore and a hot equilibrating isochore.  

The second issue is related to the preservation of system-bath correlations during the cycle. Consider for instance the second cycle mentioned above, with a cold interrupted isochore. After the hot equilibrating isochore, $S$ and $B_h$ can be strongly correlated when the interaction strength is large. Such correlations are not destroyed by the unitary expansion stroke, and in general only partially affected by the interrupted cold isochore. Then, at the end of the cycle, correlations between $S$ and $B_h$ will  still be present. These residual correlations can play an important role in the switching on work $W_{\rm on}^h$ since now $S$ and $B_h$ are correlated. In the cycle analyzed in the previous section, we assumed that the correlations between $S$ and $B_h$ were entirely destroyed which was legitimized by the fact that $S$ reaches equilibrium with $B_c$. However, interrupting the cold isochore can result in significant residual correlations. Keeping track of these system-bath correlations over several cycles would require a full non-equilibrium treatment akin to Refs. \cite{Newman_2020,Wiedmann_2020,Pezzutto_2016}, and since these effects are in principle non-negligible, we consider that the hot equilibrating isochore is a {\it weak} coupling isochore. This yields vanishing correlations between $S$ and $B_h$ and thus avoids the issue of keeping track of correlations between strokes. With these two points in mind, we briefly present the cycles designed for interrupted isochores, beginning with the cycle composed of a weak, equilibrating, cold isochore and an interrupted hot isochore, represented in Fig. \ref{fig:4}(a).

\subsection{Interrupted cycle analysis}

 {\it First stroke--}. The initial state of $S$ is $\rho_1:=\rho^{c, \rm th}_S$, produced by the weak interaction with $B_c$, and assuming $\tau_c$ is long enough so that  equilibration takes place. Apart from the initial state, the first stroke is the same as before with $\rho_2 = U_{\rm comp}\rho^{c, \rm th}_S U_{\rm comp}^{\dag}$. Hence the corresponding work expenditure is 
$W_1:={\rm Tr} [\rho_2H_h-\rho_1H_c] =  p_c^{\rm th} (\omega_h -\omega_c)\geq0$.

{\it Second stroke--}. For the hot isochore the switching on work contribution again vanishes, but now the isochore is stopped at an arbitrary time $\tau_h$. The composite state of $S+B_h$ at this instant of time is denoted by $\rho_{SB_h}(\tau_h)$ and the corresponding reduced state of $S$ is $\rho^h_S(\tau_h)={\rm Tr}_{B_h}[\rho_{SB_h}(\tau_h)]$, with $\rho_3=\rho^h_S(\tau_h)$. The associated exchanged heat is 
 \bea
 Q^{(h)}_h &:= -&{\rm Tr}\left[(\rho_{SB_h}(\tau_h) - \rho_2\rho_{B_h}^{\rm th})H_{B_h}\right] \nn\\
 &=&  {\rm Tr}[(\rho_3 - \rho_2)H_h] + {\rm Tr}[\rho_{SB_h}(\tau_h)H_{I_h}]\nn\\
  &=&  \omega_h[p_h(\tau_h)-p_c^{\rm th}] - W_{\rm off}^h(\tau_h)
 \eea
where $p_h(\tau_h) := \bra +|\rho^h_S(\tau_h)|+\ket$.
  
 {\it Third stroke--}. The expansion stroke takes the system Hamiltonian $H_h$ of the hot adiabat back to $H_c$, but the initial state is now $\rho^h_S(\tau_h)$, such that $\rho_4 = U_{\rm exp}\rho^h_S(\tau_h)U_{\rm exp}^{\dag}$. The work associated with this third stroke is then given as
 \be
 W_3= {\rm Tr}[ \rho_4 H_c - \rho^h_S(\tau_h) H_h] = p_h(\tau_h)(\omega_c - \omega_h)\leq0.
 \ee
 
 {\it Fourth stroke--.} The last stroke is the weak, cold isochore, maintained for a time $\tau_c$ such that $S$ reaches an equilibrium state $\rho^{\rm th}_S$, thereby closing the cycle. The associated heat exchange is 
 \bea
 Q^{(h)}_c &=& {\rm Tr} [(\rho_{SB_c}(\tau_c) -\rho_4\rho_{B_c}^{\rm th})H_c] + {\rm Tr}[\rho_{SB_c}(\tau_c)H_{I_c}]\nn\\
 &=& \omega_c[p^{\rm th}_c-p_h(\tau_h)] - W_{\rm off}^c
 \eea
with $W_{\rm off}^c:= -{\rm Tr}[\rho^{\rm ss}_{SB_c}H_{I_c}]$ close to zero since the coupling between $S$ and $B_c$ is weak.

\begin{figure}[t!]
\centering
\includegraphics[scale=0.55]{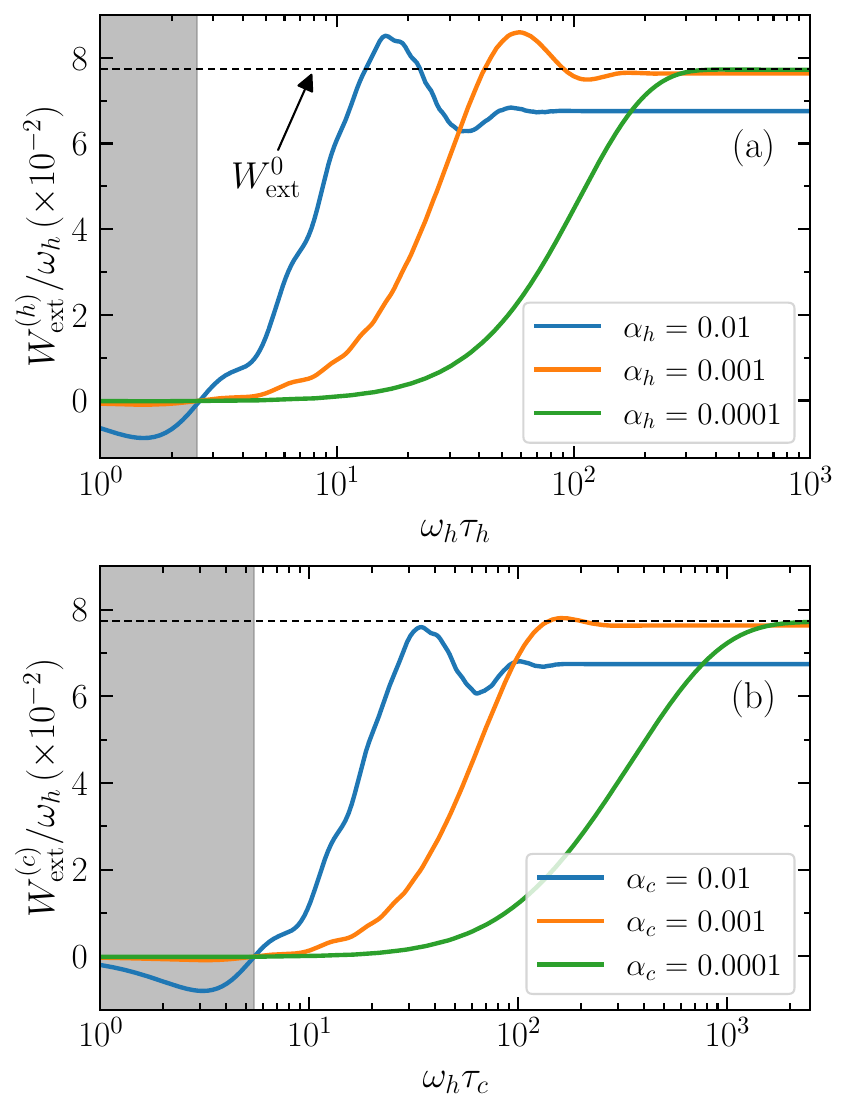}
\caption{Interrupted isochores: extracted work per cycle $W^{(j)}_{\rm ext}$ against the time duration of the (a) interrupted hot isochore, and (b) interrupted cold isochore for coupling strengths; $\alpha_j = 10^{-2}\omega_h$ (blue solid line), $\alpha_j = 10^{-3}\omega_h$ (orange solid line), and $\alpha_j = 10^{-4}\omega_h$ (green solid line). The black dashed line corresponds to the work extracted in the vanishing coupling limit, $W_{\rm ext}^0$. The filled regions indicate times for which the cycle produces a negative net work output for all $\alpha_j$.}
\label{fig:5}
\end{figure}

Similar to the previous section, the corresponding efficiency and output power are
\be
\eta^{(h)} := \frac{W^{(h)}_{\rm ext}}{Q^{(h)}_h}  = \eta_{\rm Otto} - \frac{\omega_c}{\omega_h}\frac{W_{\rm off}^c/\omega_c + W_{\rm off}^h(\tau_h)/\omega_h}{p_h(\tau_h)-p^{\rm th}_c - W_{\rm off}^h(\tau_h)/\omega_h},
\ee
and
\be
{\cal P}^{(h)} = \frac{W^{(h)}_{\rm ext}}{\tau_{\rm cyc}} \simeq \frac{W^{(h)}_{\rm ext}}{\tau_h + \tau^{\rm eq}_c}.
\ee
Finally, the alternate cycle with an interrupted cold isochore and weak equilibrating hot isochore is obtained analogously to the above, see Fig. \ref{fig:4}(b). In particular, the corresponding efficiency reads 
\be\label{eq:effcoldiso}
\eta^{(c)} := \frac{W^{(c)}_{\rm ext}}{Q^{(c)}_h}  = \eta_{\rm Otto} - \frac{\omega_c}{\omega_h}\frac{W_{\rm off}^c(\tau_c)/\omega_c + W_{\rm off}^h/\omega_h}{p^{\rm th}_h-p_c^{+}(\tau_c) - W_{\rm off}^h/\omega_h}.
\ee
In what follows we perform all simulations for the interrupted cycles using the same parameters given in Sec. \ref{sec:2}. The coupling strength of the weak, equilibriating isochore is also set to be $\alpha^{{\rm eq}}_j=10^{-4}\omega_h$ in both cases.

\begin{figure}[t]
\centering
\includegraphics[scale=0.55]{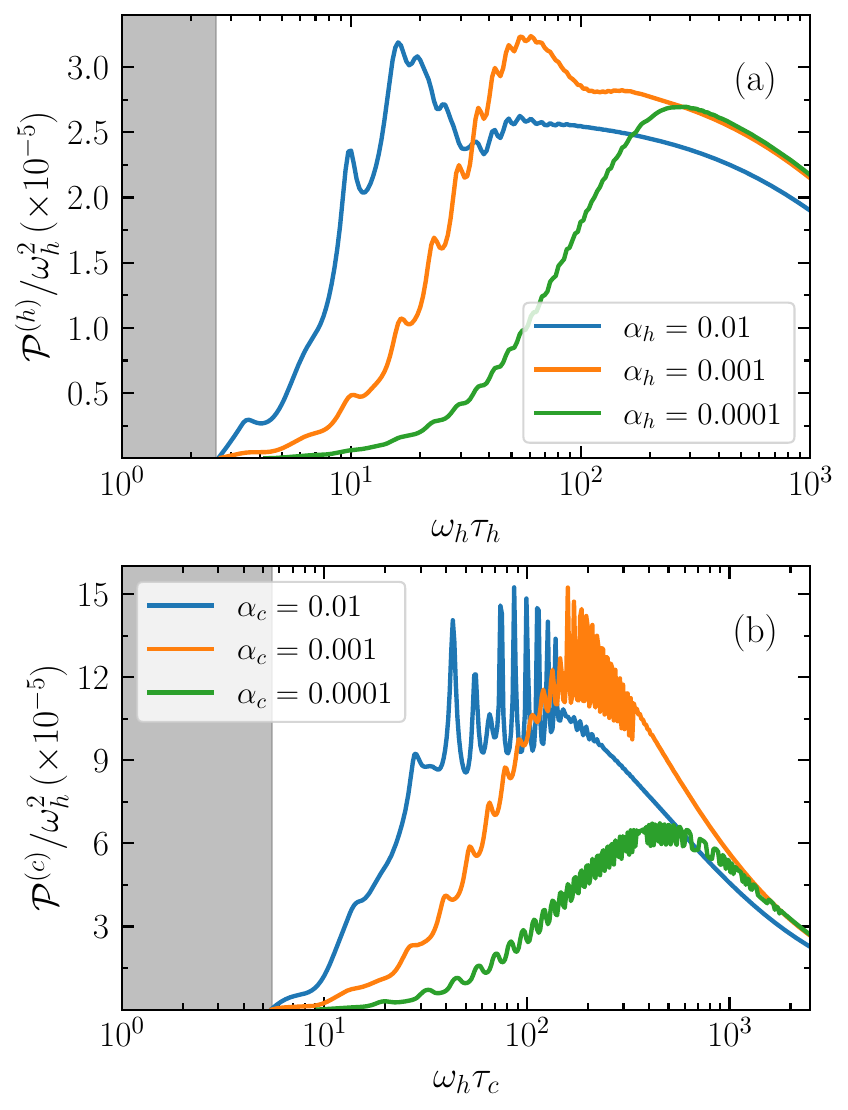}
\caption{Interrupted isochores: Power output $\mathcal{P}^{(j)}$ of the cycle against the time duration of the (a) interrupted hot isochore, and (b) interrupted cold isochore for the same three coupling strengths $\alpha_j$ depicted in Fig. \ref{fig:5}.}
\label{fig:6}
\end{figure}

\begin{figure}[t]
\centering
\includegraphics[scale=0.55]{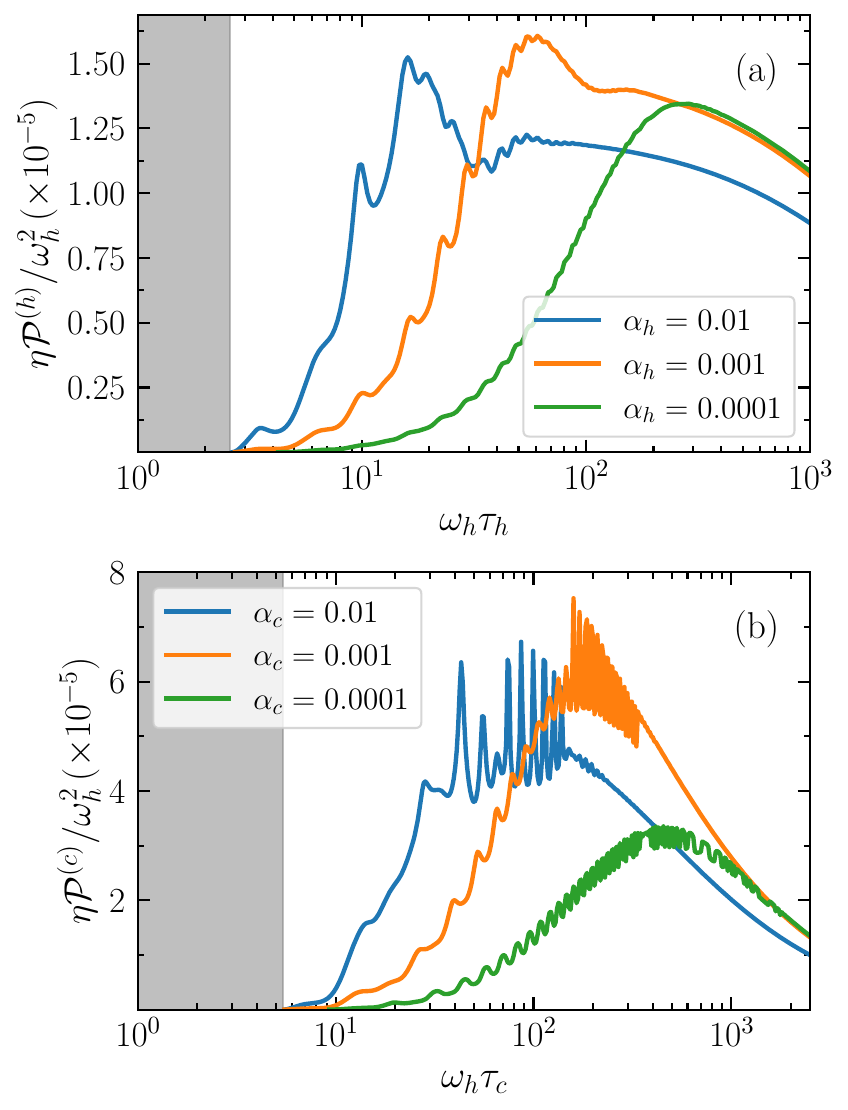}
\caption{Interrupted isochores: HFOM $\eta\mathcal{P}^{(j)}$ plotted against the time duration of the (a) interrupted hot isochore, and (b) interrupted cold isochore for the same three coupling strengths $\alpha_j$ depicted in Fig. \ref{fig:5}.}
\label{fig:7}
\end{figure}

\subsection{Results}

Figure \ref{fig:5} presents plots of the extracted work per cycle $W_\text{ext}$ against the time durations of the interrupted hot and cold isochores for various coupling strengths $\alpha_j$. For short isochore durations $\tau_j\ll \tau^{\rm eq}_j$, we find the net work output to be negative for each version of the cycle. This is due to the switching off cost $W^h_{\text{off}}+W^c_{\text{off}}$ at these times being larger than the work extracted along the expansion stroke $-(W_1+W_3)$ (not shown). On the other hand, by increasing the duration of the isochoric strokes the net work becomes positive, and above certain times for $\alpha_j=10^{-2}\omega_h$ and $\alpha_j=10^{-3}\omega_h$ yields a work output larger than that of the equivalent uninterrupted cycle (for $\alpha_j=10^{-4}\omega_h$, the work only increases monotonically until its saturates close to $W^0_{\rm ext}$). Remarkably, with an interrupted hot isochore, we find that it is even possible to extract more work than what is \textit{maximally achievable at vanishing coupling}. This is a direct consequence of the over-equilibration phenomenon observed in Fig. \ref{fig:3}(b): more specifically, since the excited population $p_h(\tau_h)$ is larger when the hot isochore is stopped before equilibration, the TLS is initialized at a higher effective temperature immediately before the application of the expansion stroke. As this corresponds to more heat having been absorbed by the TLS, it follows that more energy can be extracted from the system in the form of work. 

In parallel, for an interrupted cold isochore the TLS reaches a lower effective temperature such that more heat is dissipated into the cold bath, which leads to less work having to be invested during the compression stroke. The work increase associated with this enhanced cooling effect, however, is seen to be less significant in comparison to the analogous heating effect of the interrupted hot isochore, with only a very slight advantage gained over the weak coupling cycle for $\alpha_c=10^{-3}\omega_h$. 

Following these results, one might interpret the possibility of surpassing the vanishing coupling work output $W^0_{\rm ext}$ as a way to also reach an efficiency higher than the Otto efficiency. However, surpassing this limit is more subtle than just increasing the total extracted work: in Appendix \ref{appbeyondC} we demonstrate that this may only occur when the interrupted isochores are more reversible than those implemented at vanishing coupling, which in practice implies $\eta^{(j)}\leq \eta_{\rm Otto}$. In the same appendix this result is corroborated against numeric calculations of $\eta^{(j)}$. At strong to moderate bath coupling $\alpha_j=10^{-2}\omega_h$ and $\alpha_j=10^{-3}\omega_h$, we also find $\eta^{(j)}$ to attain larger values than for a weak interrupted isochore ($\alpha_j=10^{-4}\omega_h$) for certain stopping times $\tau_j$ below $\tau^{\rm eq}_j$. However, as $\tau_j$ approaches $\tau^{\rm eq}_j$ the weak coupling efficiency eventually surpasses that for $\alpha_j=10^{-2}\omega_h$ and $\alpha_j=10^{-3}\omega_h$, and equilibrates at a value closer to $\eta_{\rm Otto}$. 

\begin{figure*}[t]
\centering
\includegraphics[scale=0.405]{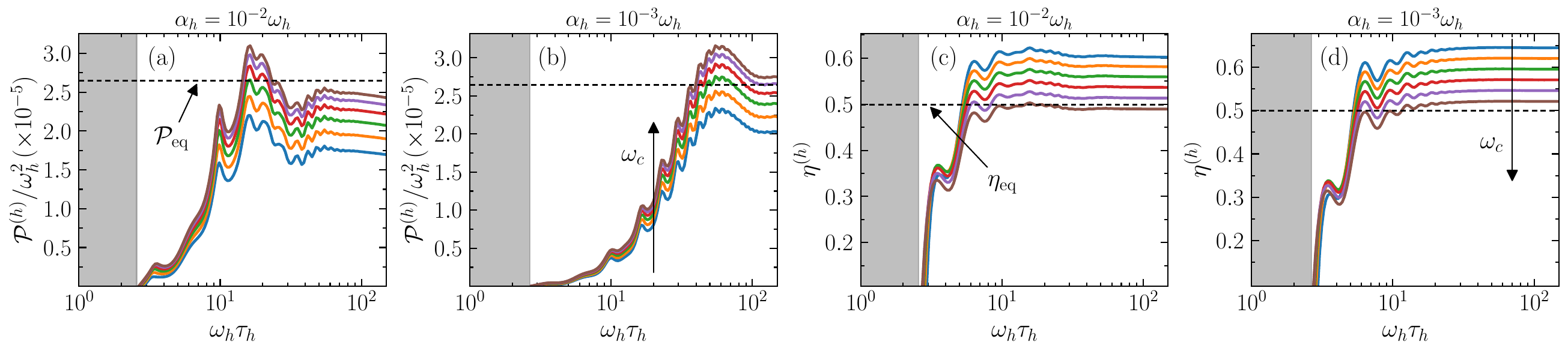}\\
\caption{Comparison of the output power $\mathcal{P}^{(h)}$ and efficiency $\eta^{(h)}$ for an interrupted hot isochore with $\omega_c/\omega_h=\{0.35, 0.375, 0.4, 0.425, 0.45, 0.475\}$ (blue solid line to brown solid line): Panels (a), (b) show the power against the duration time of the hot isochore for coupling strengths $\alpha_h=10^{-2}\omega_h$ and $\alpha_h=10^{-3}\omega_h$. Panels (c), (d) show the efficiency for the same coupling strength $\alpha_h$. In all cases the coupling to the cold bath is taken to be weak, $\alpha^{\rm eq}_c=10^{-4}\omega_h$. On the lefthand side the black dashed line indicates the power output $\mathcal{P}_{\rm eq}$ of the weak coupling cycle $\alpha_h=10^{-4}\omega_h$ with equilibrating isochores and $\omega_c = 0.5\omega_h$, while on the righthand side it corresponds to the efficiency $\eta_{\rm eq}\approx\eta_{\rm Otto}$ of the same cycle. The filled region indicates times for which the net work output of the interrupted cycle is negative for all $\omega_c$-values.}
\label{fig:8}
\end{figure*}

In Figs. \ref{fig:6} and \ref{fig:7} we show the behavior of the corresponding output power $\mathcal{P}^{(j)}$ and HFOM $\eta\mathcal{P}^{(j)}$. To calculate $\mathcal{P}^{(j)}$ and $\eta\mathcal{P}^{(j)}$ the equilibriation time $\tau^{\rm eq}_j$ of the weakly coupled isochore is determined as before through integrating the HEOM (\ref{eq:heom}) until convergence to the steady state is reached (see Appendix \ref{app:numeqtime}). Interestingly, we observe here the possibility of increasing the power output of the cycle by interrupting either one of the isochores. The over-equilibration phenomenon effective at finite (strong) coupling mainly contributes to this power increase. Note that this enhancement of the power is also reflected in the HFOM. Thus, it entails that strong coupling effects are beneficial to the performance of the interrupted engine despite the accompanying decrease in efficiency $\eta^{(j)}$ relative to the weak coupling cycle with equilibrating isochores. 

\begin{figure*}[t]
\centering
\includegraphics[scale=0.405]{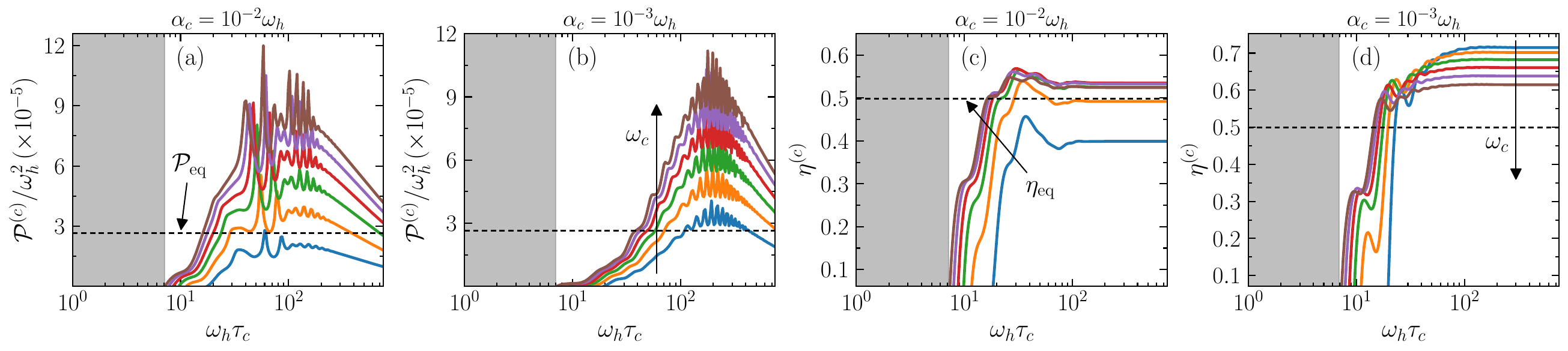}\\
\caption{Comparison of the output power $\mathcal{P}^{(c)}$ and efficiency $\eta^{(c)}$ for an interrupted cold isochore with $\omega_c/\omega_h=\{0.25, 0.275, 0.3, 0.325, 0.35, 0.375\}$ (blue solid line to brown solid line): Panels (a), (b) show the power against the duration time of the cold isochore for coupling strengths $\alpha_c=10^{-2}\omega_h$ and $\alpha_c=10^{-3}\omega_h$. Panels (c), (d) show the efficiency for the same coupling strength $\alpha_c$. In all cases the coupling to the hot bath is taken to be weak, $\alpha^{\rm eq}_h=10^{-4}\omega_h$. On the lefthand side the black dashed line indicates the power output $\mathcal{P}_{\rm eq}$ of the weak coupling cycle $\alpha_c=10^{-4}\omega_h$ with equilibrating isochores and $\omega_c = 0.5 \omega_h$, while on the righthand side it corresponds to the efficiency $\eta_{\rm eq}\approx\eta_{\rm Otto}$ of the same cycle. The filled region indicates times for which the net work output of the interrupted cycle is negative for all $\omega_c$-values.}
\label{fig:9}
\end{figure*}

Having established that an increase in the output power is possible for interrupted cycles at a slight cost to the efficiency, we now proceed to examine how this gain in power may be ``converted" to gains in efficiency. This is illustrated in Figs. \ref{fig:8} and \ref{fig:9} by tuning the value of the system frequency $\omega_c$ along the cold adiabat. More precisely, in Figs. \ref{fig:8}(a)-(b), the power $\mathcal{P}^{(h)}$ is shown against the duration of the interrupted hot isochore for several values of $\omega_c$. The corresponding efficiencies $\eta^{(h)}$ are plotted in Fig. \ref{fig:8}(c) and (d). Here, the black dashed lines represent the value of the power ${\cal P}_\text{eq}$ and efficiency $\eta_\text{eq}$ of the equivalent weak coupling cycle with equilibrating isochores for $\omega_c = 0.5 \omega_h$, which corresponds to the value of $\omega_c$ used in all previous figures. One can see in Fig. \ref{fig:8}(a) that for $\omega_c = 0.425\omega_h$, given by the red curve, the maximum power slightly surpasses the weak coupling power ${\cal P}_\text{eq}$, while the corresponding efficiency in Fig. \ref{fig:8}(c) significantly surpasses the weak coupling one, $\eta_\text{eq}$. This trend is even more pronounced in Figs. \ref{fig:8}(b) and (d) for $\omega_c = 0.4\omega_h$.       

Figure \ref{fig:9} presents similar plots for an interrupted cold isochore where the gains in efficiency $\eta^{(c)}$ are even larger compared to Fig. \ref{fig:8}(d). There is, however, a notable drop in $\eta^{(c)}$ at strong coupling $\alpha_c=10^{-2}\omega_h$ when the system frequency is increased beyond $\omega_c=0.325\omega_h$, contrary to the behavior seen in Figs. \ref{fig:8}(d) and Fig. \ref{fig:9}(c)-(d). The reason for this comes from the negative work contributions to the efficiency in Eq. (\ref{eq:effcoldiso}): these contributions, stemming mainly from $W_{\rm off}^c$, do increase with decreasing $\omega_c$, since in the denominator the population $p_c(\tau_c)$ increases with decreasing $\omega_c$. As a result, the efficiency of the cycle is also lowered. This effect does not occur for $\alpha_c = 10^{-3} \omega_h$ because at this coupling strength the corrections on the right-hand side of Eq. (\ref{eq:effcoldiso}) become negligible.

The overall conclusion from Figs. \ref{fig:8} and \ref{fig:9} is that, at fixed power, strong interactions combined with interrupted isochores can yield a larger efficiency compared to a standard weak coupling cycle. Conversely, one can also design strongly coupled cycles with the same efficiency as a standard weak coupling one but with larger output power, or even strongly coupled cycles with simultaneously both larger efficiency and output power.

\section{Conclusions}\label{sec:4}
In summary, we have shown that strong coupling effects can enhance the performance of quantum Otto engines; both in terms of the output power and a HFOM accounting for the trade-off between power and efficiency. Additionally, at fixed output power, strong coupling Otto cycles have been found to reach higher efficiencies compared to their weak coupling counterparts. In particular, these enhancements are achieved by combining strong coupling with interrupting isochores before reaching full equilibration.

The phenomenon of over-equilibration, for which the populations temporarily exceed their steady state value before converging to it, contributes significantly to the magnitude of these enhancements. This raises the question of whether this over-equilibration phenomenon could allow one to surpass the Otto efficiency. In Appendix \ref{appbeyondC}, we elaborate more on this question, and conclude that this would require strong interrupted isochores which are more reversible than weak coupling equilibrating isochores. Whether this could be achieved using techniques inspired from the optimal control of open systems \cite{Dann_2019} is left for future research.

It is important to keep in mind that the strong coupling enhancements do not scale with the coupling strength. Excessive coupling strength would degrade the performance until reaching zero (or even negative) work extraction. This is essentially due to the additional work costs associated with decoupling the system and baths. Alternatively, one could mitigate such costs by implementing a non-adiabatic drive during the unitary strokes so as to remove coherences \cite{OptimalCycle} generated from the interaction with the baths. The corresponding simulations are left for future studies.

Finally, more elaborate designs of quantum Otto engines could be investigated, including cases where both isochores are interrupted before reaching equilibrium \cite{Newman_2020,Wiedmann_2020,Pezzutto_2016}. In general this would require keeping track of system-bath correlations between strokes. Although avoided here, this could in principle be treated under the same HEOM framework by tracking the state of the ADOs around the full cycle, rather than only during the isochoric strokes. Hence one would be able to, for example, assess the role that system-bath correlations and entanglement have on the performance of the interrupted cycle. Additionally, extensions to Carnot cycles may be considered in line with the analysis of Ref. \cite{Perarnau_2018}.

\acknowledgements
CLL acknowledges funding from French National Research Agency (ANR) under grant ANR-20-ERC9-0010 (project QSTEAM). GP and FP acknowledge funding by the South African Quantum Technology Initiative (SA QuTI) through the Department of Science and Innovation of South Africa. CLL would like to warmly thank Patrice Camati and Jonas Floriano Gomes dos Santos for interesting discussions. 

\appendix

\begin{widetext}

\section{Upper bound for the coupling energy}\label{appupperbound}
In this section we show that the coupling energy between an arbitrary system $S$ of interest and a bath $B$, while in general negative, can take positive values. 

We consider that $S$ and $B$ are initially uncoupled, respectively in states $\rho_S(0)$ and thermal state at inverse temperature $\beta$, with local Hamiltonians $H_S^0$ and $H_B$. At some instant of time $t=0$, a coupling term $H_I$ (of arbitrary strength) is switched on, and the ensemble $S+B$ is left to evolve for an arbitrary time $t$. What is the sign of the coupling energy $E_\text{int}(t) := {\rm Tr}[\rho_{SB}(t)H_I]$ at that instant of time $t$? And if one waits until equilibrium (when $S+B$ reaches the mean force Gibbs state), what is the sign of $E_\text{int}^\text{eq} := {\rm Tr}[\rho_{SB}^\text{eq}H_I]$?
To answer these questions, let us consider the following protocol.
For arbitrary initial state $\rho_S(0)$, one can always perform a quench $H_S^0 \rightarrow \tilde H_S^0$ (unitary drive faster than the evolution of the system) such that $\rho_S(0)$ is a thermal state of $\tilde H_S^0$ at inverse temperature $\beta$. The Hamiltonian $\tilde H_S^0$ can be constructed as follows. One first expresses $\rho_S^0$ in its diagonal form,
\be
\rho_S^0 := \sum_{l=1}^L r_l |r_l \ket\bra r_l|,
\ee
and
\be
H_S^0 = \sum_{k=1}^K e_k |e_k \ket\bra e_k|,
\ee 
with $1\leq L \leq K \leq +\infty$. For $l \in [1;L]$, we introduce pseudo-energies $E_l$ as 
\be
E_l := -\frac{1}{\beta}(\ln r_l + c),
\ee
 where $c$ is a free constant one can use to choose the origin of energy. Then, if $\rho_S^0$ is full rank, meaning if $L=K$, we define $\tilde H_S^0$ as 
 \be
 \tilde  H_S^0 := \sum_{l=1}^L E_L |r_l \ket\bra r_l|.
 \ee
However, if $\rho_S^0$ is not full rank, $L<K$, we define $\tilde  H_S^0$ as
 \be
 \tilde H_S^0 := \sum_{l=1}^L E_l |r_l\ket\bra r_l| + \sum_{l=L+1}^K E_l |e_l\ket\bra e_l|,
 \ee
 where, for all $l \geq L+1$, the pseudo-energies $E_l$ are chosen such that $E_l \gg 1/\beta$. With this choice, one can verify that 
 \be
 w_S[\beta, \tilde H_S^0] := \frac{1}{\tilde Z}e^{-\beta \tilde  H_S^0}  = \rho_S^0,
 \ee
 with $\tilde Z:= {\rm Tr} [e^{-\beta  \tilde H_S^0}] = \sum_{l=1}^K e^{-\beta E_l}$. For the full-rank situation, the equality is exact. The equality becomes approximate in the non-full-rank situation, but the approximation is exponentially good for large $E_l$, $l>L$. This concludes the construction of an Hamiltonian $H_S$ such that $\rho_S^0$ can be expressed as a thermal state at inverse temperature $\beta$ with respect to $H_S$.  
 The work involved in this unitary drive is 
 \be
 W_\text{dr} = {\rm Tr}[\rho_S^0(\tilde H_S^0-H_S^0)].
 \ee

In a second step, a weak coupling between $S$ and $B$ is switched on, and the local Hamiltonian of $S$ is driven back from $\tilde H_S^0$ to $H_S^0$, \textit{in an adiabatic quasi-static way}, so that the system $S$ has time to equilibrate to the instantaneous thermal state before any significant change occurs in its Hamiltonian. The final state of $S$ is then $\rho_S^{\rm th} = w_S[\beta, H_S^0] = Z^{-1}e^{-\beta H_S^0}$, $Z={\rm Tr}[e^{-\beta H_S^0}]$, and the work spent in this reversible quasi-static isotherm is 
\bea
W_\text{iso} &=&  \Delta F \nn\\
&=& {\rm Tr}[\rho_S^{\rm th}H_S^0] -\frac{1}{\beta}S(\rho_S^{\rm th}) - \Big({\rm Tr}[ \rho_S^0 \tilde H_S^0] - \frac{1}{\beta}S(\rho_S^0)\Big)\nn\\
&=& \frac{1}{\beta} \ln \frac{\tilde Z}{Z}; 
\eea
namely, the variation of the Helmholtz free energy. 

The third step consists of switching on the interaction $H_I$ between $S$ and $B$, which is not assumed to be weak. Note that before proceeding to this third step, one can choose to switch off the coupling between $S$ and $B$
considered in the previous step, but since it is a weak, the associated energetic cost can be neglected. Then, the coupling $H_I$ is switched on in a quasi static way, for instance $\lambda (t) H_I$, with $\lambda = 0 $ at the beginning of the process and $\lambda =1$ and the end, with slow variation of $\lambda(t)$ such that $S+B$ is in equilibrium at all time (mean force Gibbs state). Then, when $\lambda$ reaches 1, we are in the same state $\rho_{SB}^{\text eq}$ as if we switched on $H_I$ instantaneously and waited for equilibration. The crucial difference is that the protocol detailed above is fully reversible (and therefore infinitely slow in practice, thus not viable). The external work required for this step is denoted by $W_{I,{\rm on}}^\text{rev}$ and is given by the variation of the Helmholtz free energy of $S+B$.  

Finally, for the last step, after reaching $\rho_{SB}^{\rm eq}$, we switch off $H_I$ infinitely slowly, in a quasi-static way similarly to the switch on process. One ends up with the local thermal state $\rho_S^\text{th}$ and $\rho_B^\text{th}$, and Hamiltonians $H_S^0$ and $H_B$, respectively. In other words, we come back to the situation prior to the third step, and $W_{I,{\rm off}}^\text{rev}$, the work involved in the switching off process, is given by the variation of the Helmholtz free energy and therefore exactly cancels $W_{I,{\rm on}}^\text{rev}$.

Then, the total work spent from the initial situation $[\rho_S^0, H_S^0]$ to the final one $[\rho_S^\text{th}, H_S^0]$, is 
\bea
W_\text{on and off}^\text{rev} &=& W_\text{dr} + W_\text{iso}\nn\\
&=& - \frac{1}{\beta} D[\rho_S^0|\rho_S^\text{th}],
\eea
where $D[\sigma|\rho]:= {\rm Tr}[\sigma(\ln \sigma -\ln \rho)] \geq 0$ is the relative entropy.

By comparison, starting from $\rho_S^0$ and $H_S^0$, we consider arbitrary switching on of $H_I$, we then wait until equilibration $\rho_{SB}^\text{eq}$. The work spent in this switching on protocol is larger than the work spent in the reversible protocol described above because we start and end in the same state and Hamiltonian,
\be
W_\text{on} \geq W_\text{on}^\text{rev} := W_\text{dr} + W_\text{iso} + W_\text{I,on}^\text{rev}.
\ee
We can then consider an arbitrary switching off process, such that the final local Hamiltonian is $H_S^0$, followed by thermal equilibration at inverse temperature $\beta$ (vanishing coupling to bath at the same temperature). Again, the switching off work associated to this arbitrary protocol is larger than the reversible switching off protocol, which is because the initial and final states and Hamiltonians are the same,
\be
W_\text{off} \geq  W_{I,\text{off}}^\text{rev}.
\ee
Finally, we obtain that the switching on and off work for an {\it arbitrary} protocol satisfies
\be
W_\text{on and off} \geq W_\text{on and off}^\text{rev} = -\frac{1}{\beta}D[\rho_S^0|\rho_S^{\rm th}].
\ee
 
Let us now consider a particular protocol, sometimes called {\it quench}, such that the switching on and off happens on a timescale much faster than the evolution time of $S+B$. In between the switching on and off quenches, we wait for equilibration of $S+B$. Hence,
\bea
W_\text{on}^\text{quench} &=& {\rm Tr}[\rho_S^0\rho_B^{\rm th} H_I],\\
W_\text{off}^\text{quench} &=& - {\rm Tr}[\rho_{SB}^{\rm eq} H_I].
\eea
If it is additionally assumed that ${\rm Tr}[\rho_B^{\rm th}H_I]=0$, then we have 
\be
-{\rm Tr}[\rho_{SB}^\text{eq} H_I] \geq -\frac{1}{\beta} D[\rho_S^0|\rho_S^{\rm th}],
\ee
which provides an interesting upper bound for the coupling energy. In particular, if $S$ is initially in a thermal state at inverse temperature $\beta$, then $\rho_S^0 = \rho_S^\text{th}$ and $W_\text{on and off}^\text{rev} = 0$, implying,
\bea
W_\text{on and off} \geq 0,
\eea
for an arbitrary protocol and  
\be
{\rm Tr}[\rho_{SB}^\text{eq}H_I] \leq 0,
\ee
when applied to a quench protocol. Since $\rho_{SB}^\text{eq}$ does not depend on the initial state, one concludes that, at equilibrium, we always have 
\be
{\rm Tr}[\rho_{SB}^\text{eq}H_I] \leq 0.
\ee
In particular, this only holds if ${\rm Tr}[\rho_S^{\rm th}\rho_B^{\rm th} H_I]=0$, otherwise it is replaced by the more general inequality ${\rm Tr}[\rho_{SB}^\text{eq}H_I] \leq {\rm Tr}[\rho_S^{\rm th}\rho_B^{\rm th} H_I]$. \\

Conversely, if $\rho_S^0 \ne \rho_S^{\rm th}$, and we consider a quench switching on and off at arbitrary time $t$, the overall work is lower bounded by the reversible one since the two protocols start in the same situation and both end in the same situation $\rho_S^{\rm th}$ and $H_S^0$ (assuming a work-free, vanishingly weak coupling to the thermal bath to bring $S$ back to $\rho_S^{\rm th}$). Consequently, we have
\be
-{\rm Tr}[\rho_{SB}(t)H_I] = W_{\rm on}^\text{quench} + W_{\rm off}^\text{quench}(t) \geq -\frac{1}{\beta}D[\rho_S^0|\rho_S^{\rm th}].
\ee
This allows for the possibility of having 
\be
{\rm Tr}[\rho_{SB}(t)H_I] \geq 0,
\ee
as well as in general for an arbitrary protocol starting from non-thermal state,
\be
W_\text{on and off} \leq 0.
\ee
In summary, we have shown that we always have ${\rm Tr}[\rho_{SB}^\text{eq}H_I] \leq {\rm Tr}[\rho_S^{\rm th}\rho_B^{\rm th} H_I]$, which becomes ${\rm Tr}[\rho_{SB}^\text{eq}H_I] \leq 0$ when ${\rm Tr}[\rho_S^{\rm th}\rho_B^{\rm th} H_I] = 0$. Additionally, provided $\rho_S^0 \ne \rho_S^{\rm th}$, there is a possibility that the switching on and off ``quench" has a negative work cost (meaning work is extracted from the switching on and off), and in particular ${\rm Tr}[\rho_{SB}(t)H_I] \geq 0$.  \\

\section{Hierarchical equations of motion} \label{appHEOM}
Within the constraints outlined in the main text, the reduced system density matrix in the interaction picture admits an exact solution of the form $\tilde{\rho}_S(t)=\mathcal{T}[\mathcal{F}[V^{\pm}_j(t),C_j(t)]\rho_S(0)]$, where 
\bea\label{eq:fvif}
    \mathcal{F}[V^{\pm}_j(t),C_j(t)] = \exp\left(-\int^t_{0}d\tau\int^{\tau}_{0}ds\,V^-_j(\tau)\big[C^+_j(\tau-s)V^-_j(s) + iC^-_j(\tau-s)V^+_j(s)\big]\right)
\eea
is the Feynman-Vernon influence functional of the $j$th bath \cite{Feynman_1963}, $C^{\pm}_j(t)$ are the symmetric $+$ (real) and anti-symmetric $-$ (imaginary) parts of $C_j(t)$, and $\mathcal{T}$ is the chronological time-ordering operator (by convention we use $t=0$ to denote the start time of the system-bath interaction along each isochore). The superoperators $V^{\pm}_j(t)$ are defined as 
\bea
	V^-_j\rho = [V_j,\rho],\qquad V^+_j\rho = \{V_j,\rho\}, 
\eea
with $V_j(t)$ the system coupling operator in the interaction picture. Formally, the above expression for $\mathcal{F}[V^{\pm}_j(t),C_j(t)]$ can be derived via a second-order cumulant expansion of the full system-bath density matrix $\tilde{\rho}_{SB_j}(t)$ \cite{FrancescoBook}.

To construct the HEOM we first assume that 
\be
    C^{\pm}_j(t) = \sum^{K^{\pm}_j}_{k=0}c^{\pm}_{jk}e^{-iz^{\pm}_{jk}t}, \qquad t\geq0,
\ee
where the coefficients $c^{\pm}_{jk}$ and $z^{\pm}_{jk}$ may in general be complex. The integral in Eq. (\ref{eq:fvif}) can then be formally solved in the Schr\"{o}dinger picture with the introduction of the ADOs: 
\begin{align}\label{eq:ADO}
	\hat{\rho}_{\vec{n}_j}(t) &= \mathcal{T}\bigg\{\exp\bigg[\int^t_0d\tau\mathcal{L}_S(\tau)\bigg]\bigg\}\mathcal{T}\Bigg\{\prod^{K^+_j}_{k=0}\left[-ic^+_{jk}\int^t_0d\tau\,e^{-iz^+_{jk}(t-\tau)}V^-_j(\tau)\right]^{n_{+k}}\nn\\
	&\times\prod^{K^-_j}_{k=0}\left[c^-_{jk}\int^t_0d\tau\,e^{-iz^-_{jk}(t-\tau)}V^+_j(\tau)\right]^{n_{-k}}\mathcal{F}[V^{\pm}_j(t),C_j(t)]\Bigg\}\rho_S(0). 
\end{align}
Here, $\mathcal{L}_S(t)\rho=-i[H_S(t),\rho]$ is the system Liouvillian, and $\vec{n}_j=(n_{+1},...,n_{+K^+_j};n_{-1},...,n_{-K^-_j})^T$ is a multi-index column vector labelling each of the ADOs of the $j$th bath with indices $n_{\pm k}\geq0$. The ADO for which all indices are zero $n_{\pm k}=0$ corresponds to the reduced system density matrix $\rho_S(t)=\rho_{(0,...,0)}(t)$. By taking the time derivative of Eq. (\ref{eq:ADO}) and applying the chain rule, the equation of motion for each ADO is obtained as
\begin{align}\label{eq:HEOM}
	\frac{\partial}{\partial t}\hat{\rho}_{\vec{n}_j}(t) &= \Bigg(\mathcal{L}_S(t) - i\sum_{i=\{+,-\}}\sum^{K^{i}_j}_{k=0}n_{ik}z^{i}_{jk}\Bigg)\hat{\rho}_{\vec{n}_j}(t) \nonumber\\
								      & -i\sum^{K^+_j}_{k=0}c^+_{jk}n_{+k}V^-_j\hat{\rho}_{\vec{n}_j-\vec{e}_{+k}}(t) + \sum^{K^-_j}_{k=0}c^-_{jk}n_{-k}V^+_j\hat{\rho}_{\vec{n}_j-\vec{e}_{-k}}(t) -i\sum_{i=\{+,-\}}\sum^{K^{i}_j}_{k=0}V^-_j\hat{\rho}_{\vec{n}_j+\vec{e}_{ik}}(t), 
\end{align}
where terms written as $\rho_{\vec{n}_j\pm\vec{e}_{\pm k}}(t)$ reference ADOs with index $n_{\pm k}$ raised or lowered by one. In principle, the hierarchy represented by Eq. (\ref{eq:HEOM}) is infinite in depth, but can in practice be truncated to a finite number of levels by limiting the maximum index values of the ADOs to $M_j$, such that $n_{\pm k}\in\{0,...,M_j\}$. The accuracy of all numerical results is verified in the usual way by increasing the value of the parameters $M_j$ and $K^{\pm}_j$ until convergence is reached (here, we find $K^{\pm}_h=1$ and $K^{\pm}_c=2$ is sufficient to achieve convergence across all coupling strengths).

For the underdamped spectral densities $J_j(\omega)$ considered in the main text, the two-time correlation functions $C_j(t)$ may be evaluated analytically with the help of contour integration techniques to obtain 
\begin{align}\label{eq:C_contour}
C_j(t)&:= \sum^{\infty}_{k=0}c_{jk}e^{-iz_{jk}t} \nn\\
&=\frac{\alpha_j\omega^2_{0,j}}{4\Omega_j}\Bigg[\Bigg(\coth{\left(\frac{\beta_j(\Omega_j-i\gamma_j)}{2}\right)}e^{-i\Omega_j t} +\text{c.c.}\Bigg) - e^{i\Omega_jt} + e^{-i\Omega_jt}\Bigg]e^{-\gamma_jt} \nn\\
       &\qquad -\frac{4\alpha_j\omega^2_{0,j}\gamma_j}{\beta_j}\sum^{\infty}_{k=1}\frac{\nu_{jk}}{(\Omega^2_j+\gamma^2_j-\nu^2_{jk})^2+4\Omega^2_j\nu_{jk}^2}e^{-\nu_{jk}t},
\end{align}
where $\nu_{jk}=\frac{2\pi k}{\beta_j}$ ($k\in \mathbb{Z}^+$) are the Matsubara frequencies, $\Omega_j=\sqrt{\omega^2_{0,j}-(\Gamma_j/2)^2}$, and $\gamma_j=\Gamma_j/2$. With this expression it is easily shown that $C^-_j(t)$ is in the required exponential form, while a similar expansion for $C^+_j(t)$ can be obtained by treating the Matsubara terms above a certain cut-off $K_j$ as being singular. In particular, if $1/\nu_{jk}$ is much smaller than the time scales of interest for $k\geq K_j$ (where $K_j\gg \omega_j/\nu_{j1}$), then we may approximate Eq. (\ref{eq:C_contour}) by
\be\label{eq:C_del_fun}
	C_j(t) \approx \sum^{K_j}_{k=0}c_{jk}e^{-iz_{ik}t} + \sum^{\infty}_{k=K_j+1}\frac{c_{jk}}{\nu_{jk}}\delta(t), \qquad K_j\geq 0.
\ee
To account for the delta function component of the correlation functions one can add a renormalization term (or so-called Tanimura terminator) to the HEOM through \cite{Ishizaki_2005}
\be
	\mathcal{L}_S(t)\rightarrow\mathcal{L}_S(t) - \left(\sum^{\infty}_{k=K_j+1}\frac{c_{jk}}{\nu_{jk}}\right)V^-_jV^-_j. 
\ee
The whole Matsubara sum for $K_j=0$ may be determined analytically as
\begin{align}\label{eq:Mats_sum}
	\sum^{\infty}_{k=1}\frac{c_{jk}}{\nu_{jk}} = -\frac{2\alpha_j\gamma_j\omega^2_{0,j}}{\beta_j}\left[\frac{\beta_j\gamma_j\sinh(\beta_j\Omega_j) + \beta_j\Omega_j\sin(\beta_j\gamma_j)}{4\Omega_j\gamma_j(\Omega^2_i+\gamma^2_j)[\cosh(\beta_j\Omega_j)-\cos(\beta_j\gamma_j)]} - \frac{1}{(\Omega_j^2+\gamma^2_j)^2}\right],
\end{align}
such that the quantity $\sum^{K_j}_{k=1}(c_{jk}/\nu_{jk})$ may then be subtracted off the above to obtain the prefactor multiplying the delta function in Eq. (\ref{eq:C_del_fun}).

\section{Further details on the equilibration time} \label{app:numeqtime}

\begin{figure*}[t]
\centering
\includegraphics[scale=0.47]{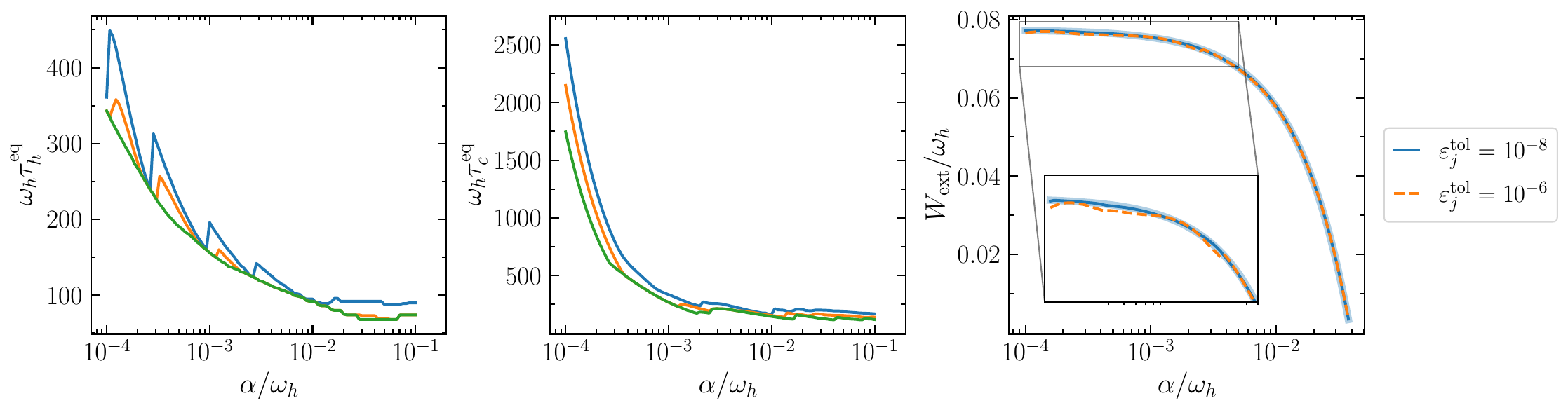}
\caption{Equilibration time versus the coupling strength $\alpha$ of (a) the hot isochore, and (b) the cold isochore, for different fidelity tolerances $\varepsilon^{\rm tol}_j=\{10^{-8}, 10^{-7}, 10^{-6}\}$ (blue, orange and green solid lines). Panel (c) shows the net output work determined from exact calculations of the TLS steady state (light blue solid line) and that obtained from evolving the system up to times $\tau^{\text{eq}}_{h/c}$ along the hot/cold isochores with a given tolerance $\varepsilon^{\rm tol}_j=10^{-6}$ (orange dashed line) and $\varepsilon^{\rm tol}_j=10^{-8}$ (blue solid line).}
\label{fig:10}
\end{figure*}

As mentioned in the main text, the equilibration time $\tau^{\rm eq}_j$ of the system along each isochore is computed by integrating the HEOM until convergence to the steady state is realized. To implement this in practice we first determine the steady state $\rho^{(h/c)}_S$ corresponding to hot/cold bath interaction using the in-built steady state solver of the Python package QuTiP \cite{Johansson_2013,Lambert_2023}. We then iteratively propagate the solution of the HEOM [Eq. (\ref{eq:HEOM})] up to a fixed time $t\geq0$, at which point the reduced system state $\rho_S(t)$ is sufficiently close to $\rho^{j}_S$, so that
\be
    \tau^{\rm eq}_j = \max\{t\in\mathbb{R}^+\,|\,\,1-F(\rho_S(t),\rho^{j}_S)\geq \varepsilon^{\rm tol}_j\},
\ee 
where $F(\rho,\sigma)={\rm Tr}[\sqrt{\rho^{1/2}\sigma\rho^{1/2}}]$ denotes the fidelity between the two states $\rho$ and $\sigma$, and $\varepsilon^{\rm tol}_j$ is a specified tolerance threshold. 

Following this procedure, the equilibration times $\tau^{\rm eq}_h$ and $\tau^{\rm eq}_c$ of the strong coupling Otto cycle are displayed in Figs. \ref{fig:10}(a) and (b) for various choices of tolerance thresholds. For all numerical results of the output power and HFOM presented in the main text we use an equal fidelity tolerance of $\varepsilon^{\rm tol}_j=10^{-8}$ for $j=h,c$. As one might expect, the equilibration times are found to generally increase by lowering the tolerance threshold, although there are notable differences in how $\tau^{\rm eq}_{j}$ scales with the coupling strength $\alpha$ of each isochore. For the cold isochore, the equilibration time scales relatively smoothly in the coupling as $\tau^{\rm eq}_c \sim O(1/\alpha)$, except with a small number of ``jumps" occurring in $\tau^{\rm eq}_c$ at certain values of the coupling below $\alpha=10^{-3}\omega_h$. The hot isochore, on the other hand, exhibits a greater sensitivity to changes in $\alpha$ for tolerance values $\varepsilon^{\rm tol}_h\leq 10^{-7}$, where jumps in the equilibration time are more frequent and occur at more regular intervals in the coupling. However, by increasing $\varepsilon^{\rm tol}_h$ the larger jumps subside and we retrieve a smoother scaling similar to that of the cold isochore. 

Based on this difference one might question whether these jumps in $\tau^{\rm tol}_h$, which only appear at certain tolerances, are intrinsic to the evolution or simply a numerical artefact of the procedure. As a consistency check, in Fig. \ref{fig:10}(c) we plot the work output of the cycle by evolving the system up to the corresponding equilibration time, and compare this with the work obtained from calculating the steady state numerically (i.e., without directly evolving the system state). Here we find that the equilibration times obtained for $\varepsilon^{\rm tol}_j=10^{-8}$ reproduce with high fidelity the work computed from the steady state, whereas those obtained with a lower tolerance $\varepsilon^{\rm tol}_j=10^{-6}$ show noticeably larger discrepancies at weaker couplings. Hence, this would suggest that a tolerance of $10^{-8}$ (or lower) is sufficient to obtain an accurate estimate of the equilibration times of the two isochores.

\section{Beyond Otto efficiency?}\label{appbeyondC}
We asked in the main text whether the ``over-equilibration" could be used to beat the Otto efficiency. In this section we analyse this question in full generality, without specifying whether the populations reached at the end of the isochores are beyond their equilibration values or not. One can conclude that the determinant aspect to surpass the Otto efficiency is whether work is spent or extracted during the switching on and off processes. Ultimately, and more generally, it appears that the only way to beat the Otto efficiency is to operate closer to reversibility than traditional Otto cycle, recovering the well-known statement that limitations to efficiency stems from irreversibility of the cycle. The question of whether interrupted strong isochores can be more reversible than the traditional isochores used in weak coupling Otto cycle is still an open question. In the following, we detail how one can derive such conclusions.

\begin{figure*}[t]
\includegraphics[scale=0.61]{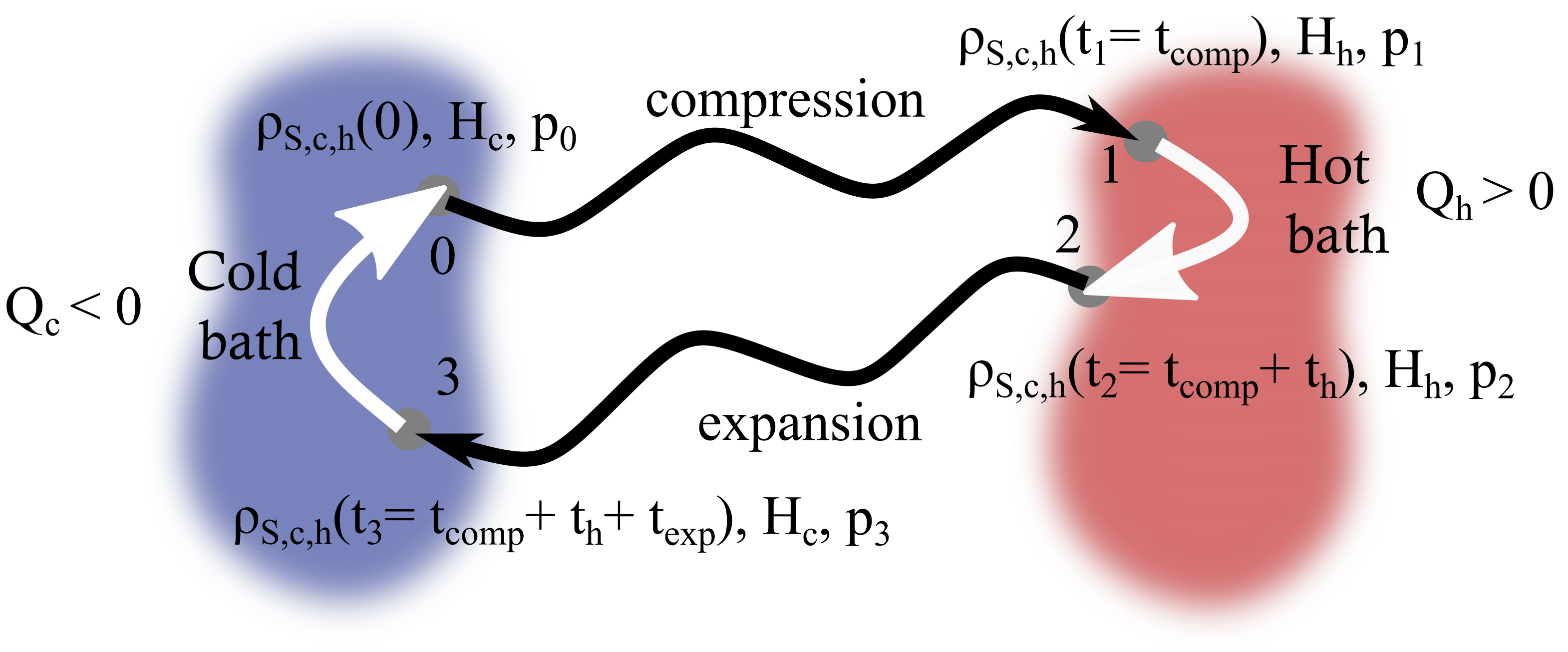}\\
\caption{Sketch of a general Otto cycle, where no assumption is made neither on the strength of the bath coupling, nor on the duration of the isochores, thus represented by thick and short white arrows, nor on the kind of driving during the unitary strokes, represented by wavy and thick black arrows. }
\label{fig:11}
\end{figure*}

\subsection{Generalized Otto cycle}

We consider a generalized Otto cycle, represented in Fig. \ref{fig:11}, where all strokes are represented by joined unitary evolution onto the global system $S$, cold bath $c$, and hot bath $h$. We again assume the working medium to be a two-level system. Accordingly, the cycle is described by the following steps:
\begin{itemize}
\item {\it Compression stroke}, from $t_0=0$ to $t_1:=t_\text{comp}$:\\ 
Arbitrary $H_S(t)$ with the restriction $H_S(0) = \omega_c\Pi^+_c:=H_c$, and $H_S(t_\text{comp})=\omega_h\Pi^+_h:=H_h$, where $\Pi^+_{h/c}=|+_{h/c}\ket\bra+_{h/c}|$ are system projectors, and $|+_h\ket$ ($|+_c\ket$) denotes the upper eigenstate of the system Hamiltonian $H_h$ ($H_c$). Starting from $\rho_{S,c,h}(0)$, the state at the end of the compression stroke is $\rho_{S,c,h}(t_1) = U_\text{comp}\rho_{S,c,h}(0)U_\text{comp}^{\dag}$. The excited populations of the system at the beginning and end of the compression stroke are $p_0:={\rm Tr}[(\Pi^+_c\otimes\mathbb{I}_{c,h})\rho_{S,c,h}(0)]$ and $p_1:={\rm Tr}[(\Pi^+_h\otimes\mathbb{I}_{c,h})\rho_{S,c,h}(t_1)]$, respectively.

\item {\it Hot isochore}, from $t_1$ to $t_2:=t_\text{comp}+t_\text{h}$: \\
The system $S$ is instantaneously coupled to the hot bath $h$ through an interaction $H_{I_h}$, at an arbitrary coupling strength $\alpha_h$. The global state at the end of the hot isochore (potentially stopped before reaching equilibrium) is $\rho_{S,c,h}(t_2) = U_h\rho_{S,c,h}(t_1)U_h^{\dag}$. The corresponding excited population is $p_2:={\rm Tr}[(\Pi^+_h\otimes\mathbb{I}_{c,h})\rho_{S,c,h}(t_2)]$. At the end of the interaction $S$ is instantaneously decoupled from $h$.

\item {\it Expansion stroke}, from $t_2$ to $t_3:=t_\text{comp}+t_\text{h}+t_\text{exp}$:\\
Arbitrary $H_S(t)$ with the restriction $H_S(t_2) = H_h$, and $H_S(t_3) = H_c$. The state at the end of the expansion stroke is $\rho_{S,c,h}(t_3) = U_\text{exp}\rho_{S,c,h}(t_2)U_\text{exp}^{\dag}$. The corresponding excited population is $p_3:={\rm Tr}[(\Pi^+_c\otimes\mathbb{I}_{c,h})\rho_{S,c,h}(t_3)]$.

\item {\it Cold isochore}, from $t_3$ to $t_4:=t_\text{comp}+t_\text{h}+t_\text{exp} + t_\text{c}$:\\
The system $S$ is instantaneously coupled to the cold bath $c$ through an interaction $H_{I_c}$, at an arbitrary coupling strength $\alpha_c$. The global state at the end of the cold isochore (potentially stopped before reaching equilibrium) is $\rho_{S,c,h}(t_4) = U_c\rho_{S,c,h}(t_3)U_c^{\dag}$. Furthermore, we assume that $\rho_{S,c,h}(0)$ as well as all other parameters (coupling strengths, stroke durations) are chosen such that $\rho_{S,c,h}(t_4) = \rho_{S,c,h}(0)$. Thus, the corresponding excited popuation is $p_4:={\rm Tr}[(\Pi^+_c\otimes\mathbb{I}_{c,h})\rho_{S,c,h}(t_4)] = p_0$. In practice, this is guaranteed by taking at least one of the two isochores long enough so that the equilibrium state is reached. Going beyond that and considering two non-equilibrating isochores would be challenging in practice, but here we assume so while still having $\rho_{S,c,h}(t_4) = \rho_{S,c,h}(0)$. Similar to the hot isochore, $S$ is instantaneously decoupled from $c$ at the end of the interaction.
\end{itemize}

Without imposing any further assumptions, we can now express all the relevant quantities determining the performance of the cycle:
\bea
W_\text{comp} &:=& \omega_h p_1 - \omega _c p_0, \\
W_{\rm on}^h &:=& \int_{t_1^-}^{t_1^+} du\,{\rm Tr}[\rho_{S,c,h}(u)\dot H_{I_h}(u)], \\
W_{\rm off}^h &:=& \int_{t_2^-}^{t_2^+} du\,{\rm Tr}[\rho_{S,c,h}(u)\dot H_{I_h}(u)], \\
W_{\text{on and off}}^h &:=& W_{\rm on}^h + W_{\rm off}^h \nn\\
&=& \int_{t_1}^{t_2} du\,{\rm Tr}[\rho_{S,c,h}(u)\dot H_{I_h}(u)], \nn\\
Q_h &:=& \omega_h(p_2 - p_1) + W_{\text{on and off}}^h = - \Delta E_h, \\
W_\text{exp} &:=& \omega_c p_3 - \omega_h p_2, \\
W_{\rm on}^c &:=& \int_{t_3^-}^{t_3^+} du\,{\rm Tr}[\rho_{S,c,h}(u)\dot H_{I_c}(u)], \\
W_{\rm off}^c &:=& \int_{t_4^-}^{t_4^+} du\,{\rm Tr}[\rho_{S,c,h}(u)\dot H_{I_c}(u)], \\
W_{\text{on and off}}^c &:=& W_{\rm on}^c + W_{\rm off}^c, \nn\\
&=& \int_{t_3}^{t_4} du\,{\rm Tr}[\rho_{S,c,h}(u)\dot H_{I_c}(u)], \nn\\
Q_c &:=&  \omega_c(p_4 - p_3) + W_{\text{on and off}}^c.
\eea
This leads to
\bea
W_\text{ext} &:=& -\left( W_\text{comp} + W_\text{exp} + W_{\rm on}^h +W_{\rm off}^h +W_{\rm on}^c+W_{\rm off}^c \right)\nn\\
0 &=&Q_c + Q_h + W_\text{ext},
\eea
where the last identity is valid assuming $p_4=p_0$. Then, the efficiency is 
\bea
\eta &:=& \frac{W_\text{ext}}{Q_h} = 1 - \frac{-Q_c}{Q_h} \nn\\
&=& 1- \frac{\omega_c(p_3-p_4) - W_{\text{on and off}}^c}{Q_h} \nn\\
&=& 1- \frac{\omega_c}{\omega_h} \nn\\
&\times&\left(1 + \frac{\omega_h(p_1-p_0 + p_3-p_2)- W_{\text{on and off}}^h - \frac{\omega_h}{\omega_c}W_{\text{on and off}}^c}{Q_h}\right).
\eea
Since $Q_h >0$ is positive by hypothesis (we are designing an engine), the efficiency is larger than the Otto efficiency $\eta_{\rm Otto} = 1 - \omega_c/\omega_h$ only if
\be
\frac{W_{\text{on and off}}^h}{\omega_h} + \frac{W_{\text{on and off}}^c}{\omega_c} + p_0-p_1 + p_2-p_3 \geq 0.
\ee
The positivity of the first two terms means that work is extracted from the total switching on and off operations on both baths. According to the previous section, this could be possible for interrupted strong isochores. The last terms, $ p_0-p_1 + p_2-p_3$, are positive when the populations decrease along the compression and expansion strokes. This occurs only if the state at the beginning of these strokes contains coherences, which can happen only for strongly coupled isochores. Thus, it seems that for two interrupted strong isochores, it could be possible to go beyond the Otto efficiency. However, as already commented in the previous section, the difficulty with two interrupted isochores is to be able to close the cycle ($p_4=p_0$, and even $\rho_{S,c,h}(t_4) = \rho_{S,c,h}(t_0)$). 

\begin{figure*}[t!]
\centering
\includegraphics[scale=0.55]{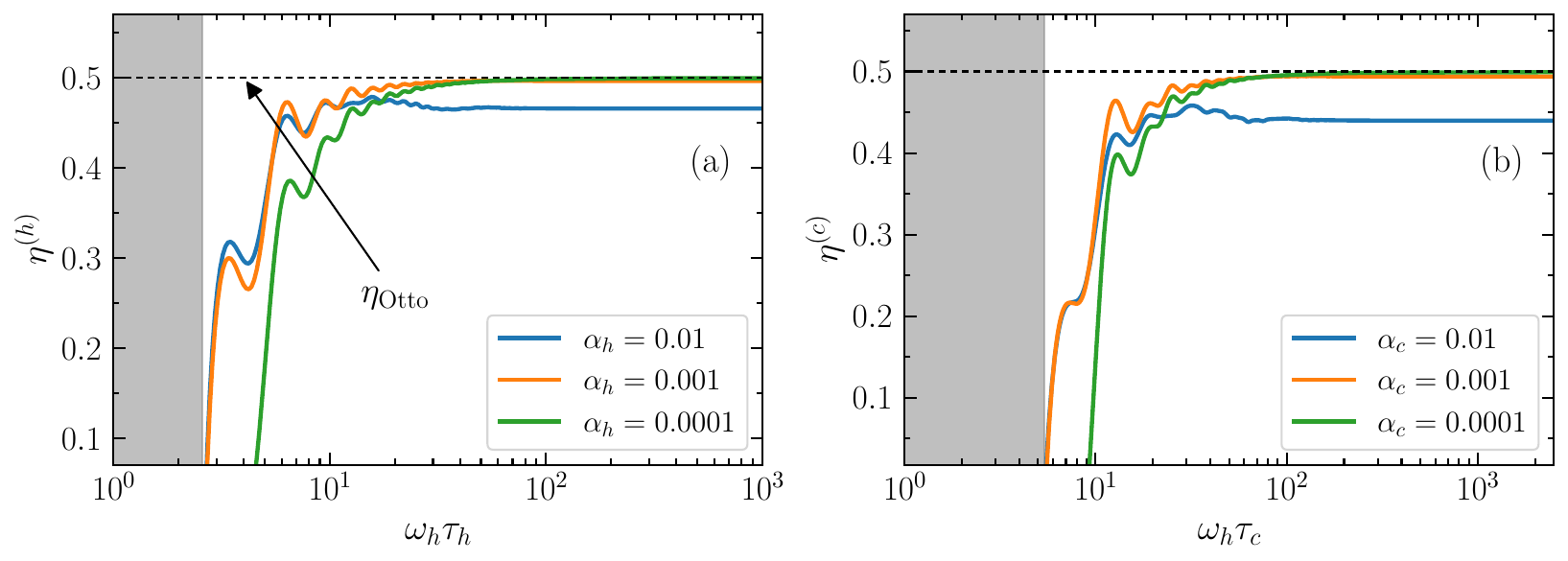}
\caption{Interrupted isochores: efficiency $\eta^{(j)}$ plotted against the time duration of the (a) interrupted hot isochore, and (b) interrupted cold isochore for coupling strengths; $\alpha_j = 10^{-2}\omega_h$ (blue solid line), $\alpha_j = 10^{-3}\omega_h$ (orange solid line), and $\alpha_j = 10^{-4}\omega_h$ (green solid line), where the black dashed line corresponds to the Otto efficiency $\eta_{\rm Otto}$. The filled regions indicate times for which the cycle produces a negative net work output for all $\alpha_j$.}
\label{fig:12}
\end{figure*}

Furthermore, using the framework introduced in \cite{CEandCLL}, we have
\bea
&&\frac{W_{\text{on and off}}^h}{\omega_h} + \frac{W_{\text{on and off}}^c}{\omega_c} + p_0-p_1 + p_2-p_3 \nn\\
&=& -\frac{\Delta E_h}{\omega_h} -\frac{\Delta E_c}{\omega_c} \nn\\
&=& -\frac{\Delta E_h^\text{th}}{\omega_h} -\frac{\Delta E_c^\text{th}}{\omega_c}  +\frac{W_h}{\omega_h} +\frac{W_c}{\omega_c} \nn\\
&=& \frac{\Delta S_S^h - \Delta I_h - D_h}{\omega_h\beta_h(0)} + \frac{\Delta S_S^c - \Delta I_c - D_c}{\omega_c\beta_c(0)}   +\frac{W_h}{\omega_h} +\frac{W_c}{\omega_c}\nn\\
&=&\Delta S_S^h\left( \frac{ 1}{\omega_h\beta_h(0)} - \frac{1}{\omega_c\beta_c(0)}\right)\nn\\
&&+ \frac{- \Delta I_h - D_h}{\omega_h\beta_h(0)} + \frac{- \Delta I_c - D_c}{\omega_c\beta_c(0)}   +\frac{W_h}{\omega_h} +\frac{W_c}{\omega_c},
\eea
where $E^{\rm th}_{c/h}$ denotes the ``thermal" energy of the baths \cite{CEandCLL}, $W_{h/c} := -\Delta (E_{h/c} - E_{h/c}^\text{th})$ is the work provided by the hot/cold bath, $\Delta S_S^{h/c}$ denotes the variation of entropy of $S$ during the hot/cold isochore, $\Delta I_{h/c}$ is the variation of correlations between $S$ and the hot/cold bath during the respective isochores, and finally $D_h : = D[\rho_S^{h,\rm th}(t_2)|\rho_S^{h,\rm th}(t_1)]$, $D_c : = D[\rho_S^{c,\rm th}(t_4)|\rho_S^{c,\rm th}(t_3)]$ are the ``thermal" distance increase associated with the hot/cold isochore. Finally, $\beta_h(0)$ and $\beta_c(0)$ are the initial effective temperature \cite{CEandCLL} of the hot and cold baths. In the last line, we use the fact that the entropy of $S$ is constant along the compression and expansion strokes, which yields $\Delta S_S^h = -\Delta S_S^c$.
With that, the efficiency can be re-expressed in the following form,
\bea
\eta &=& 1 - \frac{\beta_h(0)}{\beta_c(0)} \nn\\
&+& \frac{\omega_c}{\omega_h}\frac{-\frac{\Delta I_h +D_h}{\omega_h\beta_h(0)} -\frac{\Delta I_c +D_c}{\omega_c\beta_c(0)} +\frac{W_c}{\omega_c}+\frac{W_h}{\omega_h}}{Q_h/\omega_h}.
\eea
If the hot and cold baths are assumed to be initially in thermal states, $W_h$ and $W_c$ are necessarily negative (the baths cannot provide work, they can only receive work). Secondly, $\Delta I_h + D_h$ and $\Delta I_c + D_c$ correspond to the entropy production during the hot and cold isochores, respectively. The conclusion is that with this generalized Otto cycle, one can in principle have an efficiency as large as the Carnot efficiency, but only if no work is damped into the baths and if the isochores are reversible. As we know, this is never the case in practice. Regarding the previous considerations, it also means that one can go beyond the Otto efficiency only if -- at fixed heat absorbed from the hot bath -- the interrupted strong isochores can be made more reversible than the usual weak coupling isochores  (and if no work is damped in the baths). In other words, and as announced in the beginning of this section, the only way to beat the Otto efficiency is to be more reversible than the weak coupling Otto cycle.

\subsection{Efficiency results for interrupted Otto cycles}

Finally, we show in Fig. \ref{fig:12} the efficiencies $\eta^{(j)}$ of the interrupted Otto cycles considered in Sec. \ref{sec:3} of the main text. Interestingly, for certain stroke durations $\tau_j$ below the equilibration time (but above that for which work output is negative), the efficiency $\eta^{(j)}$ is higher at strong to moderate system-bath coupling ($\alpha_j=10^{-2}\omega_h$ and $\alpha_j=10^{-3}\omega_h$) than with a weak interrupted isochore ($\alpha_j=10^{-4}\omega_h$) for both versions of the cycle. Nonetheless, $\eta^{(j)}$ for all values of $\tau_j$ remains below the Otto efficiency $\eta_{\rm Otto}$: as stated above, this is a consequence of the interrupted isochores being less reversible than the isochores of a standard weak coupling cycle (i.e., the total entropy production along these strokes at finite system-bath coupling is larger than in the vanishing coupling case). Note that as the duration of the interrupted isochores approaches the equilibration time $\tau^{\rm eq}_j$ the efficiency of the weakly coupled cycle eventually overtakes that at stronger couplings, and for $\tau_j\geq \tau^{\rm eq}_j$ saturates at a value closer to the Otto efficiency. It should also be emphasized that, while these results suggest the bound $\eta^{(j)}\leq \eta_{\rm Otto}$, one cannot rule out the possibility of $\eta^{(j)}$ exceeding $\eta_{\rm Otto}$ for certain parameters based on the analysis of the preceeding section.

\end{widetext}

\end{document}